\documentclass[11pt,a4paper]{article}
\pdfoutput=1
\usepackage{jheppub}
\usepackage{amsfonts}
\usepackage{amssymb}
\usepackage{mathtools}
\usepackage{slashed}
\usepackage[colorlinks=true]{hyperref}

\begin{document}

\def\bibsection{\section*{References}}

\newcommand{\GeV}{\ensuremath{\, \mathrm{GeV}}}
\newcommand{\cm}{\ensuremath{\, \mathrm{cm}}}
\newcommand{\s}{\ensuremath{\, \mathrm{s}}}
\newcommand{\mub}{\ensuremath{{\bar{\mu}}}}
\newcommand{\nub}{\ensuremath{{\bar{\nu}}}}
\newcommand{\rhob}{\ensuremath{{\bar{\rho}}}}
\newcommand{\sigmab}{\ensuremath{{\bar{\sigma}}}}
\newcommand{\lambdab}{\ensuremath{{\bar{\lambda}}}}
\newcommand{\vev}{\ensuremath{{\langle v \rangle}}}
\newcommand{\Lag}{\mathcal{L}}
\newcommand{\chibar}{\bar{\chi}}
\newcommand{\ct}{\cos\theta}
\newcommand{\st}{\sin\theta}
\newcommand{\ca}{\cos\alpha}
\newcommand{\sa}{\sin\alpha}
\newcommand{\Sc}{\chibar\chi}
\newcommand{\PSc}{\chibar i \gamma_5 \chi}
\newcommand{\voL}[1]{\dfrac{\langle v \rangle^2}{2\Lambda #1} }

\let\oldFootnote\footnote
\newcommand\nextToken\relax

\renewcommand\footnote[1]{%
    \oldFootnote{#1}\futurelet\nextToken\isFootnote}

\newcommand\isFootnote{%
    \ifx\footnote\nextToken\textsuperscript{,}\fi}


\title{The Fermionic Dark Matter Higgs Portal: an effective field theory approach}

\author{Michael A.\ Fedderke,} \emailAdd{mfedderke@uchicago.edu}

\author{Jing-Yuan Chen,} \emailAdd{chjy@uchicago.edu}

\author{Edward W.\ Kolb} \emailAdd{Rocky.Kolb@uchicago.edu}

\author{and Lian-Tao Wang} \emailAdd{liantaow@uchicago.edu}

\affiliation{Enrico Fermi Institute and Kavli Institute for Cosmological Physics, The University of Chicago, Chicago, Illinois \ \ 60637-1433 }

\abstract{
We consider fermionic (Dirac or Majorana) cold thermal relic dark-matter coupling to standard-model particles through the effective dimension-5 Higgs portal operators $\Lambda^{-1} \ \mathcal{O}_{\text{DM}} \cdot H^\dagger H$, where $\mathcal{O}_{\text{DM}}$ is an admixture of scalar $\bar\chi\chi$ and pseudoscalar $\bar\chi i\gamma_5 \chi$ DM operators.  Utilizing the relic abundance requirement to fix the couplings, we consider direct detection and invisible Higgs width constraints, and map out the remaining allowed parameter space of dark-matter mass and the admixture of scalar and pseudoscalar couplings. We emphasize a subtlety which has not previously been carefully studied in the context of the EFT approach, in which an effect arising due to electroweak symmetry breaking can cause a na\"ively pure pseudoscalar coupling to induce a scalar coupling at higher order, which has important implications for direct detection bounds. We provide some comments on indirect detection bounds and collider searches.}

\date{\today}

\arxivnumber{1404.2283}

\maketitle

\section{Introduction \label{sec:introduction}}

The existence of dark matter (DM) provides solid evidence for new physics beyond the Standard Model (SM). Among the menagerie of possible dark-matter candidates that have been proposed and explored in the literature, the weakly interacting massive particle (WIMP) scenario stands out as the most compelling. In this scenario, WIMPs are established in local thermodynamic equilibrium (LTE) in the early universe through the coupling of the WIMPs to SM particles.  The present abundance of WIMPs is determined by the freeze-out from LTE of the WIMPs.  If the correct relic abundance is attained, freeze-out occurs when the temperature of the universe drops below the mass of the WIMP by a factor of 20 or so.

Within the WIMP paradigm, there are typically multiple complementary experimental probes that utilize the WIMP-SM coupling to probe the WIMP hypothesis. There are two approaches in specifying the WIMP-SM coupling.  In a top-down approach one imagines a complete enveloping model or theory that contains a WIMP and a prescription for how the WIMP couples to SM particles.  An example of this approach is the assumption of low-energy supersymmetry where the WIMP is the lightest supersymmetric particle \cite{Jungman:1995df}. The other approach is a bottom-up effective field theory (EFT) parameterization.  In the latter approach one usually assumes a DM-SM interaction of the form $\Lambda^{-n} \  \mathcal{O}_{\text{DM}} \cdot \mathcal{O}_{\text{SM}}$, where $\Lambda$ is the EFT mass scale, $\mathcal{O}_{\text{DM}}$ and $\mathcal{O}_{\text{SM}}$ are DM and SM operators that are singlets under the standard-model gauge groups \cite{Beltran:2008xg,Beltran:2010ww,Goodman:2010yf}.  An advantage of the bottom-up approach is that it provides the simplest approach to combining the different experimental approaches for WIMP discovery.   

In the EFT approach it is necessary to make assumptions for the form of $\mathcal{O}_{\text{DM}}$ and $\mathcal{O}_{\text{SM}}$.  The Higgs bilinear, $H^\dagger H$, is the lowest mass-dimension gauge-invariant operator consisting of SM matter fields. It is therefore natural to consider DM couplings to the SM via the so-called Higgs portal operators, of the form ${\mathcal{O}}_{\rm DM} \cdot H^\dagger H$. As the Higgs field plays the central role in electroweak symmetry breaking (EWSB), the Higgs field will have important effects on the dark matter mass and couplings in this scenario. There exists an extensive literature on Higgs portal dark matter; for example, refs.\ \cite{Burgess:2000yq,Patt:2006fw,Kim:2006af,Barger:2007im,
Kim:2008pp,Kanemura:2010sh,Djouadi:2011aa,Low:2011kp,Englert:2011aa,Batell:2011pz,
Fox:2011pm,Pospelov:2011yp,Englert:2011lq,LopezHonorez:2012kv,Kamenik:2012hn,Tsai:2013bt,
Carpenter:2013xra,Esch:2013rta,Fairbairn:2013uta,Greljo:2013wja,Petrov:2013nia,
Walker:2013hka,Crivellin:2014qxa,deSimone:2014pda}.

In this paper, we present a complete study of the lowest-dimensional Higgs portal coupling of fermionic dark matter. We perform a detailed study of the dark matter masses and couplings which pass current experimental bounds. Although previous work (e.g., refs.~\cite{LopezHonorez:2012kv,deSimone:2014pda}) have explored some aspects of this scenario, we extend this work by considering in a systematic fashion simultaneous contributions from both the CP-conserving (${\mathcal{O}}_{\rm DM} \sim \bar{\chi}\chi$) and CP-violating (${\mathcal{O}}_{\rm DM} \sim \bar{\chi}i\gamma_5\chi$) Higgs portal couplings. We also emphasize a subtlety which has not previously been carefully studied in the context of the EFT approach: although the CP-violating coupling only mediates highly (momentum transfer) suppressed contributions to the direct detection process at leading order, effects arising due to EWSB can generate a significant CP-conserving coupling. Although this effect is higher order in the EFT suppression scale, the lifting of the momentum-transfer suppression can greatly enhance the direct-detection cross-section over the na\"ive expectation. We carefully take this effect into account by carrying out a consistent chiral rotation. 

In our analysis, we compute the (tree-level) dark-matter annihilation cross-section and use the cosmological dark-matter relic abundance to fix the EFT suppression scale. Numerical solution of the Boltzmann equation, including a full thermal averaging of the annihilation cross-section during the freeze-out process, is carried out to accurately capture the sizable resonance and threshold effects near $2 M \sim m_h$ and $M \sim m_W$, respectively. With the EFT suppression scale thus fixed, we find that the LUX direct detection bounds \cite{Akerib:2013tjd}, and --- for $2 M < m_h$ --- Higgs invisible decay \cite{Belanger:2013xza} and total width \cite{CMS-PAS-HIG-14-002} constraints rule out significant portions of the parameter space. In this paper, we combine all constraints and map out the remaining parameter space. We do not explicitly consider indirect detection bounds, but once the inherently continuum nature of the signals and large astrophysical uncertainties are considered, these limits are expected to be weak in comparison to the other probes.

In principle, the fermionic Higgs portal couplings we consider can also contribute to signals of dark-matter production at high-energy colliders, although the validity of the EFT approach per s\'e at high energy can be degraded by perturbative unitarity issues \cite{Busoni:2013lha,Busoni:2014ty,Buchmueller:2013dya}. This issue notwithstanding, the dominant contribution probably arises from the $h \chi \chi$ coupling induced by this operator. Although a detailed analysis of the reach is beyond the scope of this paper, we can offer some brief comments here. One possible signal would be a weak boson fusion process in which a dark matter pair is produced through an off-shell Higgs, giving rise to two forward tagging jets and missing energy. An off-shell Higgs could also be produced by gluon fusion, which when combined with an initial state radiation would lead to a mono-jet plus missing energy signal. Given the sizable SM model backgrounds, we expect the reach in both of these channels to be fairly limited. Of course, the Higgs can be on-shell if $2 M < m_h$, but this scenario is already strongly constrained by limits to Higgs invisible decay signals. In the future, we expect the Higgs invisible decay limits to continue to provide stronger limits in this regime than the collider direct search. 

The rest of the paper is organized as following. In section \ref{sec:eft}, we carry out the chiral rotation and present our parameterization of the model parameters. In section \ref{sec:sigma}, we present our analytic calculation of the annihilation cross section, and examine the validity of our truncation of the EFT expansion. Our calculation of the limits from Higgs decay, relic abundance, and direct detection are presented in section \ref{sec:width}, section \ref{sec:relic}, and section \ref{sec:direct}, respectively. Finally, we combine all the constraints and present the remaining parameter space in section \ref{sec:combined}, before concluding in section \ref{sec:conclusions}. Appendix \ref{app:selected_results} contains discussion of some selected results presented in a fashion complementary to the main text.

\section{The Effective Field Theory \label{sec:eft}}

We consider a convenient parametrization of the effective pre-EWSB mass-eigenstate Lagrangian coupling mixing scalar and pseudoscalar SM-singlet fermionic DM operators to the SM via the Higgs portal $H^\dagger H$:\footnote{Unless explicitly stated, we will consider the  DM field $\chi$ to be a Dirac fermion and point out differences for the Majorana fermion case.}\footnote{The parametrization in terms of $\theta$ and $\Lambda$ is convenient for a numerical scan of the parameter space, but we should caution the reader that the ``EFT suppression'' scale $\Lambda$ in this parametrization is only approximately the scale of new physics: the scalar (CP-conserving) and pseudoscalar (CP-violating) operators can logically have different new physics scales associated with them and this gets mixed up  in our parametrization. This issue should be borne in mind when judging issues of perturbative unitarity.\label{ft:scale_caveat}}
\begin{equation}
\Lag = \Lag_{\text{SM}} + \chibar \left( i\slashed{\partial} - M_0 \right) \chi + \Lambda^{-1} \bigg( \ct \ \bar{\chi}\chi + \st \ \bar{\chi} i\gamma_5 \chi\bigg) \ H^\dagger H \ .
\label{eq:b4ewsb}
\end{equation}

As the couplings break chiral symmetry independently of the mass term, one would expect $M_0$ to be at least of order $\Lambda$, and since we are assuming that the non-SM operators in \eqref{eq:b4ewsb} do not participate in EWSB, one also expects $M_0$ and $\Lambda$ are greater than the weak scale, although we will allow $M_0 <\vev$ in this work.

After EWSB the Higgs field develops a vacuum expectation value $\vev$ and the Higgs-field content becomes (in the unitary gauge with $\vev = 246$ GeV) 
\begin{equation}
H^\dagger H \longrightarrow \frac{\vev^2}{2} +  \vev h + \frac{h^2}{2}.
\end{equation} 
The Lagrangian then becomes
\begin{align}
\Lag &= \Lag_{\text{SM}} +  \chibar  i\slashed{\partial} \chi - \left[ M_0 \chibar \chi - \frac{\vev^2}{2\Lambda} \bigg( \ct \ \bar{\chi}\chi + \st \ \bar{\chi} i\gamma_5 \chi\bigg)\right] \nonumber\\ &\qquad + \Lambda^{-1} \bigg( \ct \ \bar{\chi}\chi + \st \ \bar{\chi} i\gamma_5 \chi\bigg)\left( \vev h + \frac{1}{2} h^2 \right).
\label{eq:unrot}
\end{align}
If we were to assume instead that the DM is Majorana, we would insert the conventional factor of $1/2$ in front of every fermionic bilinear; the subsequent analysis of the Lagrangian is then unchanged from the Dirac case, modulo possible initial or final state symmetry factors in computing amplitudes. 

If $\sin\theta\neq 0$, after EWSB it is necessary to perform a chiral rotation and field redefinition to have a properly defined field with a real mass
\begin{equation}
\chi \rightarrow \exp\left( i \gamma_5\ \alpha/2 \right) \chi \quad \Rightarrow \quad \chibar \rightarrow \chibar \exp\left(i \gamma_5\ \alpha/2 \right).
\end{equation}
Note that a chiral rotation by $\alpha=\pi$ would change the sign of the mass term in \eqref{eq:unrot} and also change the sign of the interaction terms. We can thus without loss of generality take $M_0>0$, so long as we preserve the relative signs between the mass term and the interaction terms.\footnote{In our parametrization this sign can be absorbed by a redefinition $\theta \rightarrow \theta+ \pi$ leading back to the same form. Thus, by suitable choice of the quadrant in which $\theta$ lies, the form \eqref{eq:unrot} is completely general with $M_0>0$.}

After chiral rotation and field redefinition, we demand that the coefficient of $\bar{\chi}i\gamma_5\chi$ vanish in order to go to the real mass basis; this determines the proper chiral rotation and gives the mass of the field after EWSB in terms of the Lagrangian parameters (we define the mass after EWSB, $M$, as the coefficient of $-\bar{\chi}\chi$ in the rotated field variables). The requisite rotation is:
\begin{equation}
\tan\alpha = \left[ \dfrac{\langle v\rangle^2}{2\Lambda} \st \right] \left[ M_0 - \dfrac{\langle v \rangle^2}{2\Lambda} \ct \right]^{-1}\ \!\!\!.
\label{eq:tan_alpha}
\end{equation}
This of course determines $\sin^2\alpha$ and $\cos^2\alpha$, but not the (common) sign of $\cos\alpha$ and $\sin\alpha$:
\begin{align}
\cos^2\alpha & = \frac{\left(M_0-\voL{}\ct\right)^2}{\left(M_0-\voL{}\ct\right)^2+\left(\voL{}\right)^2\sin^2\theta} \qquad \text{and} \qquad \\
\sin^2\alpha & = \frac{\left(\voL{}\right)^2\sin^2\theta}{\left(M_0-\voL{}\ct\right)^2+\left(\voL{}\right)^2\sin^2\theta} \ .
\end{align}
Using this rotation angle, the mass becomes 
\begin{equation}
M = \pm \sqrt{\left(M_0-\voL{}\ct\right)^2+\left(\voL{}\right)^2\sin^2\theta} \ .
\label{eq:EMM}
\end{equation}
The signs of $M$, $\cos\alpha$, and $\sin\alpha$ are common; we choose the common sign to be ``$+$'' for $M$, $\ca = + \sqrt{\cos^2\alpha}$, and $\sa = + \sqrt{\sin^2\alpha}$.  With this choice the Lagrangian becomes\footnote{If we had chosen the opposite signs for $M$, $\ca$, and $\sa$, we could perform a further chiral rotation by $\pi$ and field definition to recover the sign conventions in \eqref{eq:ta}.}
\begin{equation}
\Lag 
= \Lag_{\text{SM}} +  \chibar  i\slashed{\partial} \chi - \chibar M  \chi + \Lambda^{-1} \left(\vev h+\frac{1}{2}h^2\right) \bigg[ \cos\xi\ \bar{\chi}\chi + \sin\xi\ \bar{\chi}i\gamma_5\chi  \bigg],
\label{eq:ta}
\end{equation}
where we have defined $\xi = \theta + \alpha$:
\begin{equation}
\cos\xi = \frac{M_0}{M}\left[ \ct-\frac{\vev^2}{2\Lambda M_0}\right] \qquad \mathrm{and} \qquad \sin\xi = \frac{M_0}{M}\st\ . \label{eq:cos_sin_xi}
\end{equation}

For a fixed value of $\Lambda$, we note that the mapping between $(M_0,\theta)$ and $(M,\xi)$ is, given our sign conventions, bijective. However, as will be explained more fully below, our analysis scans over $(M,\xi)$ and fixes $\Lambda$ by requiring the correct DM relic density. In this way, $\Lambda = \Lambda(M,\xi)$, and the mapping back to $(M_0,\theta)$ from $(M,\xi)$ with $\Lambda=\Lambda(M,\xi)$ may not be 1-to-1 in some regions of parameter space. Put another way, if one scans over $(M_0,\theta)$ and asks for the value of $\Lambda$ required to give the correct relic density, there are regions of parameter space where two or more solutions may be possible, corresponding necessarily to physically distinct scenarios (different values of $M$ and $\xi$) in the Lagrangian relevant below the electroweak phase transition. As we are never interested in the regime where we must work with $(M_0,\theta)$ (see below), this subtlety does not enter our work further (although, see appendix \ref{app:selected_results}), but it should be borne in mind in when relating parameters of some UV completion to our results; of course, if $\Lambda$ is fixed \emph{a priori}, then this concern is not applicable.

Comparing eqs.\ (\ref{eq:ta}) and (\ref{eq:b4ewsb}), it appears that the discussion about chiral rotations to have a proper mass term could have been avoided by just substituting\footnote{This substitution preserves manifest $SU(2)_L\times U(1)_Y$ gauge invariance.} $H^\dagger H \rightarrow H^\dagger H - \frac{1}{2}\vev^2 = \vev h + \frac{1}{2}h^2$ in \eqref{eq:b4ewsb}. In the spirit of effective field theories, as we do not know the origin of the mass $M_0$ in the UV theory, one would na\"ively expect we should not care whether or not $M$ in \eqref{eq:ta} has a contribution from EWSB. However, we have learned something important because, due to the pseudoscalar interaction term, making the substitution $H^\dagger H \rightarrow H^\dagger H - \frac{1}{2}\vev^2 $ in \eqref{eq:b4ewsb} --- thereby avoiding the above discussion --- is equivalent to requiring a carefully chosen phase\footnote{The presence of both normal ($\propto \bar\chi \chi$) and axial ($\propto \bar\chi i \gamma^5 \chi$) mass terms is equivalent a complex mass term ($\Lag \supset -M' \bar\chi_L \chi_R + \text{h.c.}$) with a non-zero phase for $M'$.} of the $\chi$ mass term in the effective theory above the EWSB scale, which in turn would require some conspiracy in the UV complete theory to arrange. The opposite side of the same coin is that if we do work with the form of the Lagrangian at \eqref{eq:b4ewsb}, it is unnatural to have a pure pseudoscalar coupling after EWSB\footnote{Note that it is already clear at the level of the original Lagrangian that a vanishing scalar coupling is a not naturally stabilized situation as it is not protected by any symmetry (cf.\ the case of vanishing pseudoscalar coupling, which is protected by the overall CP-symmetry of the Lagrangian). What we have really learned additionally is that EWSB itself causes changes to the pure-pseudoscalar nature of the original coupling, already at tree-level.} ($\cos\xi=0$) because this requires $\Lambda M_0 \cos\theta = \vev^2/2$, which is an ill-motivated coincidental relationship between parameters in the effective high-energy theory (and thereby, its UV completion) and the electroweak vacuum expectation value.\footnote{We would like to thank the authors of ref.~\cite{HillSolon} for sharing an early version of their work, wherein a careful matching between our \eqref{eq:b4ewsb} and  \eqref{eq:ta} is discussed.} 

Although we perform a general parameter scan, there are a few limiting cases that are interesting to consider:
\begin{enumerate}
\item $\st=0$, $\ct = \pm 1$: This would be a pure scalar interaction before EWSB.  After EWSB the interaction term is  $\pm\ \Lambda^{-1}\ \bar{\chi}\chi\left(\vev h+h^2/2\right)$ and the mass is $M=\left|M_0\mp \vev^2/2\Lambda\right|$.  Thus, a pure scalar interaction before EWSB will remain a pure scalar interaction with no admixture of pseudoscalar interactions. However, note that the mass $M$ is in general different from $M_0$. 
\item $\ct=0$, $\st=\pm 1$: This would be a pure pseudoscalar interaction before EWSB.  After EWSB the interaction term is \[ \Lambda^{-1}\ \left[-\voL{M}\ \bar{\chi}\chi \pm \sqrt{1-\left(\dfrac{\vev^2}{2\Lambda M}\right)^2}\ \bar{\chi} i\gamma_5\chi\right]\left(\vev h+h^2/2\right), \] and in both cases $M=\sqrt{M_0^2+\left(\voL{}\right)^2}>\vev^2/2\Lambda$.  Even if the Higgs portal coupling is purely pseudoscalar in the EW-symmetric Lagrangian, after EWSB a scalar term proportional to $\vev^2/2\Lambda M$ is generated.
\item $M_0 = 0$ (or more generally, $M_0 \ll \vev^2/2\Lambda$):  In this case $M=\vev^2/2\Lambda$.  If $M_0=0$, then $\cos\xi=-1$ and $\sin\xi=0$, and the interaction term is purely scalar: $\Lag \supset -\Lambda^{-1}\left(vh+h^2/2\right)\bar{\chi}\chi$.  The chiral rotation that resulted in a real mass term transforms the interaction into a purely scalar interaction irrespective of the value of $\theta$.  The only two parameters in this limit are $M$ and $\Lambda$; one of the parameters may be set by the requirement that freeze out results in the correct relic abundance.
\end{enumerate}

Whether scalar, pseudoscalar, or a combination of both, the nature of the interactions is of great importance:  annihilation through a pure scalar interaction ($\sin\xi = 0$) is velocity suppressed, while elastic scattering of WIMPs with nucleons through a pure pseudoscalar interaction ($\cos\xi = 0$) is velocity suppressed.\footnote{Strictly speaking, the interaction is momentum-transfer suppressed, but for elastic scattering this leads to velocity suppression.} If both interactions are present, then the (non-velocity-suppressed) interaction most important for direct detection (scalar) may not be the same as the (non-velocity-suppressed) interaction most important for determining the relic abundance (pseudoscalar).

We note finally that the form of the Lagrangian in terms of the chirally rotated field variables is only appropriate to use `below' the electroweak phase transition.  We restrict ourselves to considering DM lighter than 3 TeV where direct detection constraints from LUX \cite{Akerib:2013tjd} are available, so this condition is always satisfied since such DM decouples at $T \lesssim \mathcal{O}(200)$ GeV (the freeze-out temperature $T_F \sim M/x_F$ with  $x_F \sim 20-25$ \cite{kolb1994early}). `Above' the phase transition, the unrotated form should be used in the freeze-out computation, while the rotated form would be relevant to compute all present-day low-energy observables: we do not explore this regime further in this paper.

\section{The Annihilation Cross Section \label{sec:sigma}}

\begin{figure}
\begin{center}
\includegraphics[width=.6\linewidth,keepaspectratio]{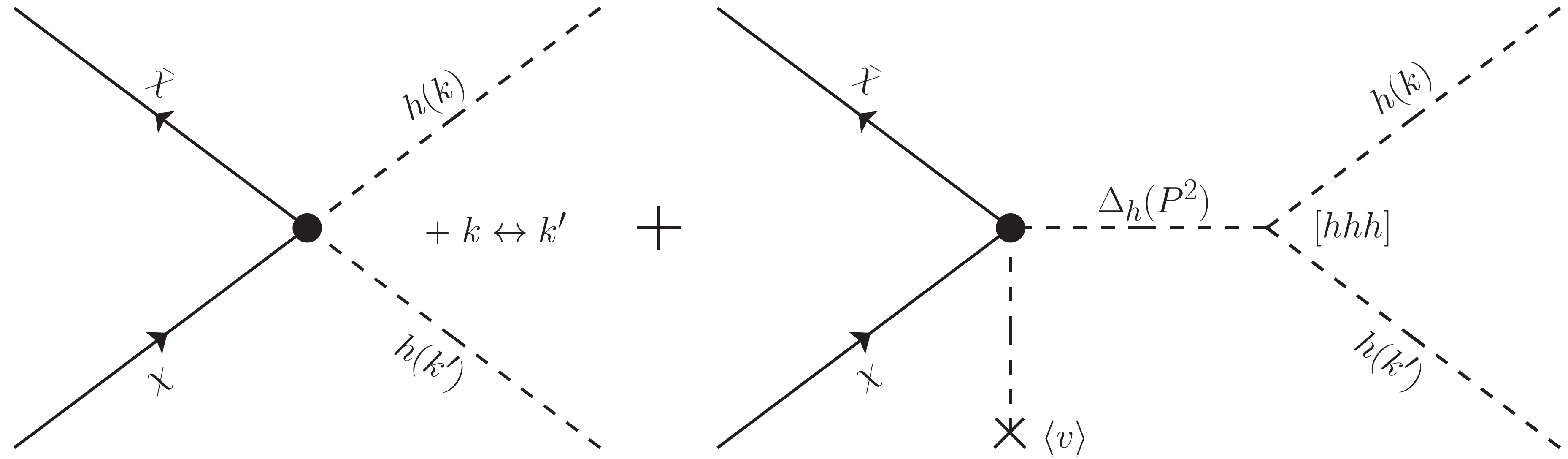} \\
\vspace*{12pt}
\includegraphics[width=.28\linewidth,keepaspectratio]{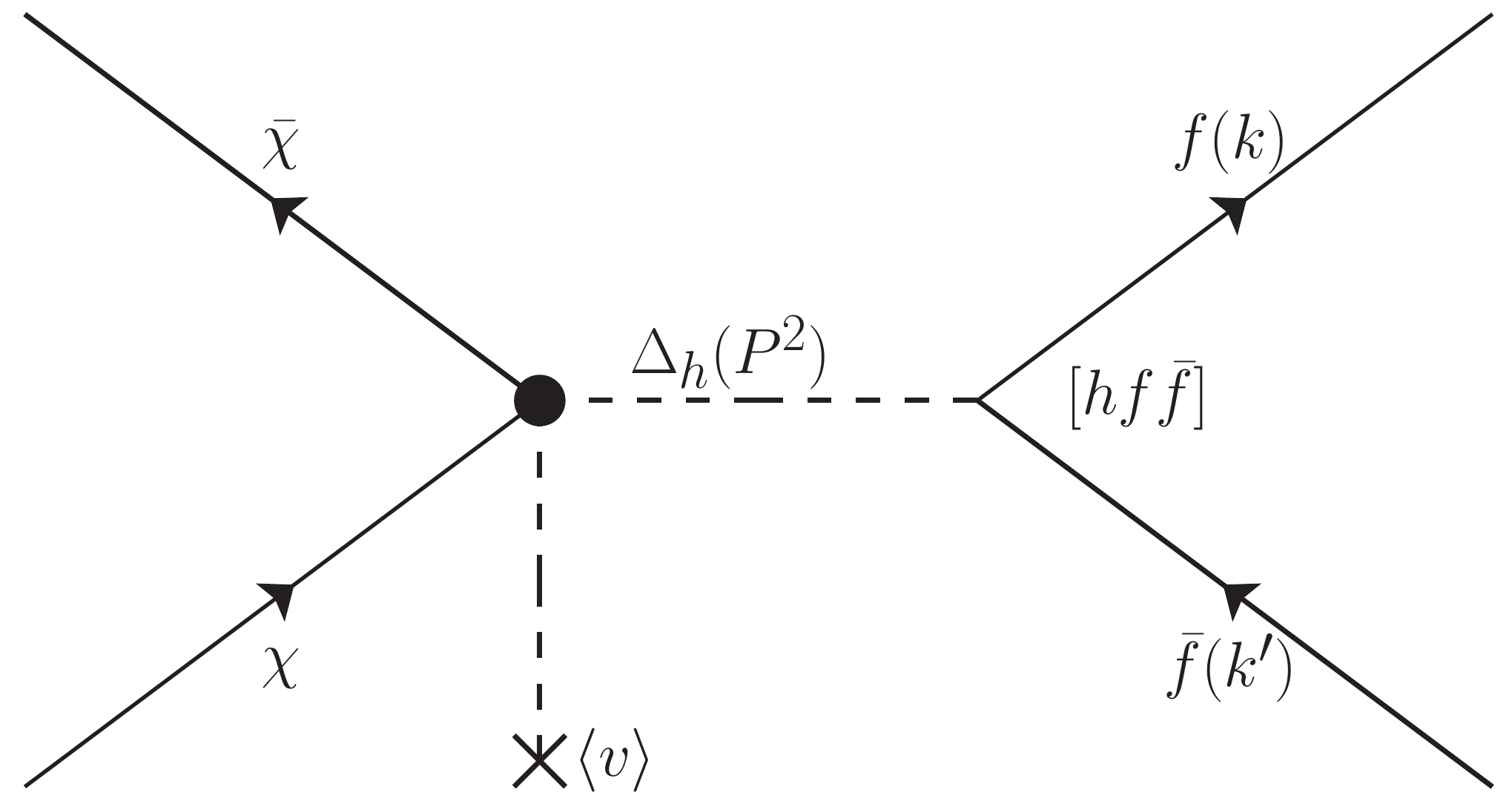} \hspace*{24pt}
\includegraphics[width=.28\linewidth,keepaspectratio]{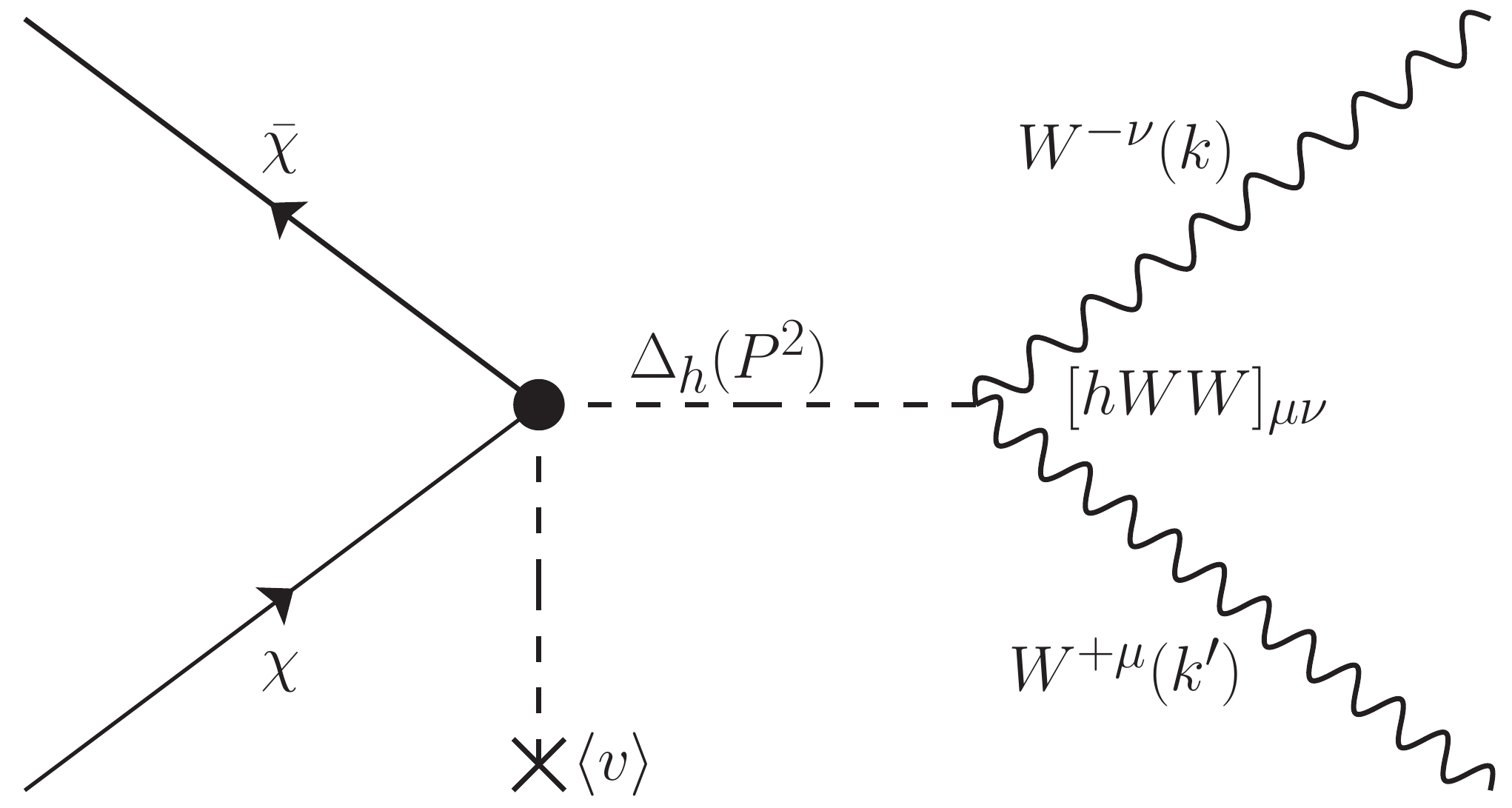} \hspace*{24pt} 
\includegraphics[width=.28\linewidth,keepaspectratio]{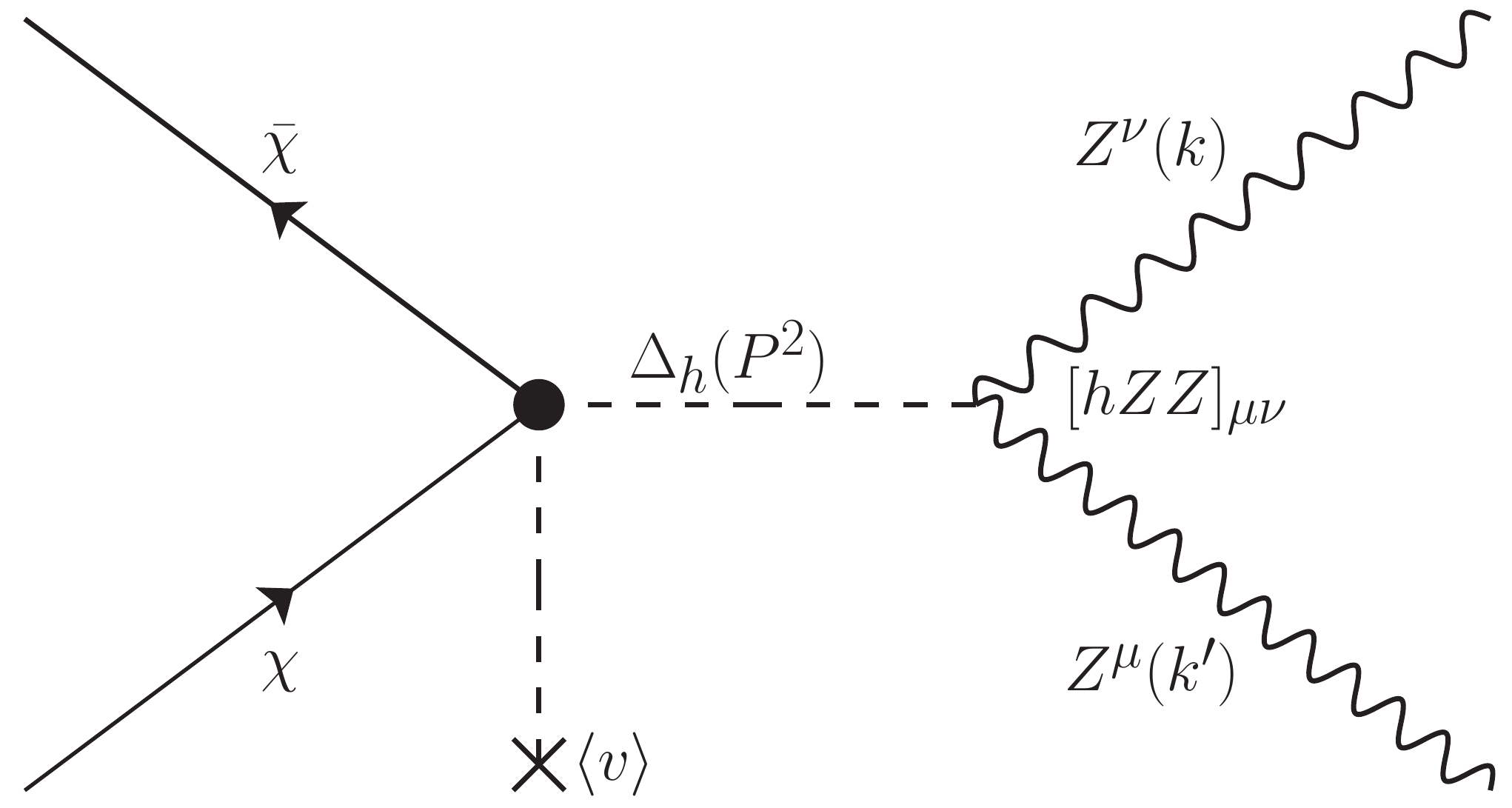}
\caption{Lowest non-vanishing order [$\mathcal{O}(\Lambda^{-1})$] final-state diagrams resulting from the Higgs-portal operator $H^\dagger H$.  \label{fig:HH}}
\end{center}
\end{figure}

We now turn the computation of the DM annihilation cross section relevant to computing the relic abundance.  It is straightforward to calculate the tree-level cross section for the $hh,\ WW,\ ZZ$ and $f\bar{f}$ 2-body final states.  In diagrammatic form, the lowest order processes are illustrated in figure \ref{fig:HH}.  The vertex functions of figure \ref{fig:HH} are \cite{Peskin:1995ev}
\begin{align}
\left[hhh\right]             &=  -3i\ \frac{m_h^2}{\vev}\ , &
\left[hf\bar{f}\right]       &=   -i\ \frac{m_f}{\vev}\ , &
\left[hWW\right]_{\mu\nu}    &=   2i\ \frac{m_W^2}{\vev}\ g_{\mu\nu}\ , &
\left[hZZ\right]_{\mu\nu}    &=   2i\ \frac{m_Z^2}{\vev}\ g_{\mu\nu}\ .
\label{eq:vertexhh}
\end{align}
The cross sections for final states $f$ may be expressed as \cite{Chen:2013gya}
\begin{equation}
\sigma_f(s;M,m') = \frac{1}{32\pi M^2} \sqrt{\frac{4M^2}{s}} \sqrt{\frac{M^2}{s-4M^2}} \sqrt{1-\frac{4m'^2}{s}} \ \ \Sigma_f(s; M, m') ,
\end{equation}
where (see also ref.\ \cite{Kim:2006af})
\begin{align}
&\Sigma_f(s; M, m') \equiv \frac{1}{4} \sum_{\text{spins}} \cdot \ \frac{1}{4\pi} 
\int d\Omega \left| \mathcal{M}_f \right|^2 \nonumber \\ 
& =   \frac{1}{4} \frac{s}{\Lambda^2} \ \dfrac{ \cos^2\xi \left( 1 - 4M^2/s \right) + \sin^2\xi } {\bigg(1-m_h^2/s\bigg)^2  + \bigg( m_h \Gamma_h /s \bigg)^2} \times \left\{ 
\begin{array}{ll}    	
  \left(1 - 4m_Z^2/s + 12m_Z^4/s^2 \right) & ZZ \\ [2ex]
2 \left(1 - 4m_W^2/s + 12m_W^4/s^2 \right) & W^+W^- \\[2ex]
   \bigg(1-4m_f^2/s\bigg) \bigg(4m_f^2/s\bigg) & f\bar{f} \\[2ex]
  \left[ \bigg(1+2m_h^2/s\bigg)^2 + \bigg( m_h\Gamma_h/s\bigg)^2 \right] & hh \ .\\  
\end{array} \right.  
\label{eq:s1}
\end{align}
In \eqref{eq:s1}, $\Gamma_h$ is the {\em total} width of the Higgs (including the partial width for $h\rightarrow \bar{\chi}\chi$ when $M<m_h/2$) and the factors of $m_f$ in the expression for $f\bar{f}$ are the running masses\footnote{To be explicit, we utilize the three-loop running masses from ref.\ \cite{Djouadi:2005gi} and references therein (e.g.\ refs.\ \cite{Vermaseren:1997fq,Chetyrkin:1997dh}).} at the scale $q^2 = s$. Note that there are no interference terms between the CP-even and CP-odd contributions to the cross-sections here; note also that we explicitly ignore the possible $3$-body and $4$-body final states mediated by one or two off-shell $W,Z$, and/or $h$, which would be important for a high-precision computation just below the thresholds for on-shell $WW,ZZ$, and/or $hh$ final states (see e.g.\ ref.\ \cite{Djouadi:2005gi} for the same point in the context of the SM Higgs branching ratio computations).

Away from resonances, the non-relativistic (NR) cross section relevant for the early-universe freeze-out calculation is obtained by the substitution $s\rightarrow 4M^2$ unless $s$ appears in the combination $s-4M^2$, in which case one substitutes $s-4M^2\rightarrow v^2M^2$, where in the NR limit $v$ is the M\o ller velocity that appears in the Boltzmann equation for the early-universe evolution of the DM density.  With these substitutions, one can see from \eqref{eq:s1} that the term proportional to $\cos^2\xi$ is proportional to $v^2$, as expected from scalar interactions.

\begin{figure}
\centering
\begin{minipage}{.45\textwidth}
  \centering
  \includegraphics[width=\textwidth,keepaspectratio]{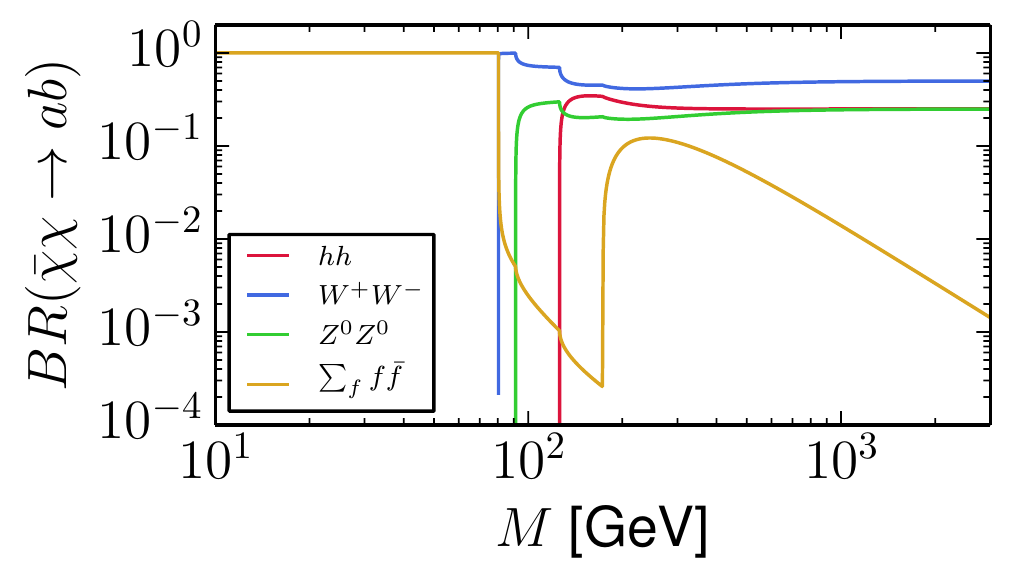}
    \caption{The branching fractions in the NR limit as a function of $M$. \label{fig:BR}}
\end{minipage}\hspace{0.3cm}%
\begin{minipage}{.45\textwidth}
  \centering
  \includegraphics[width=0.5\textwidth,keepaspectratio]{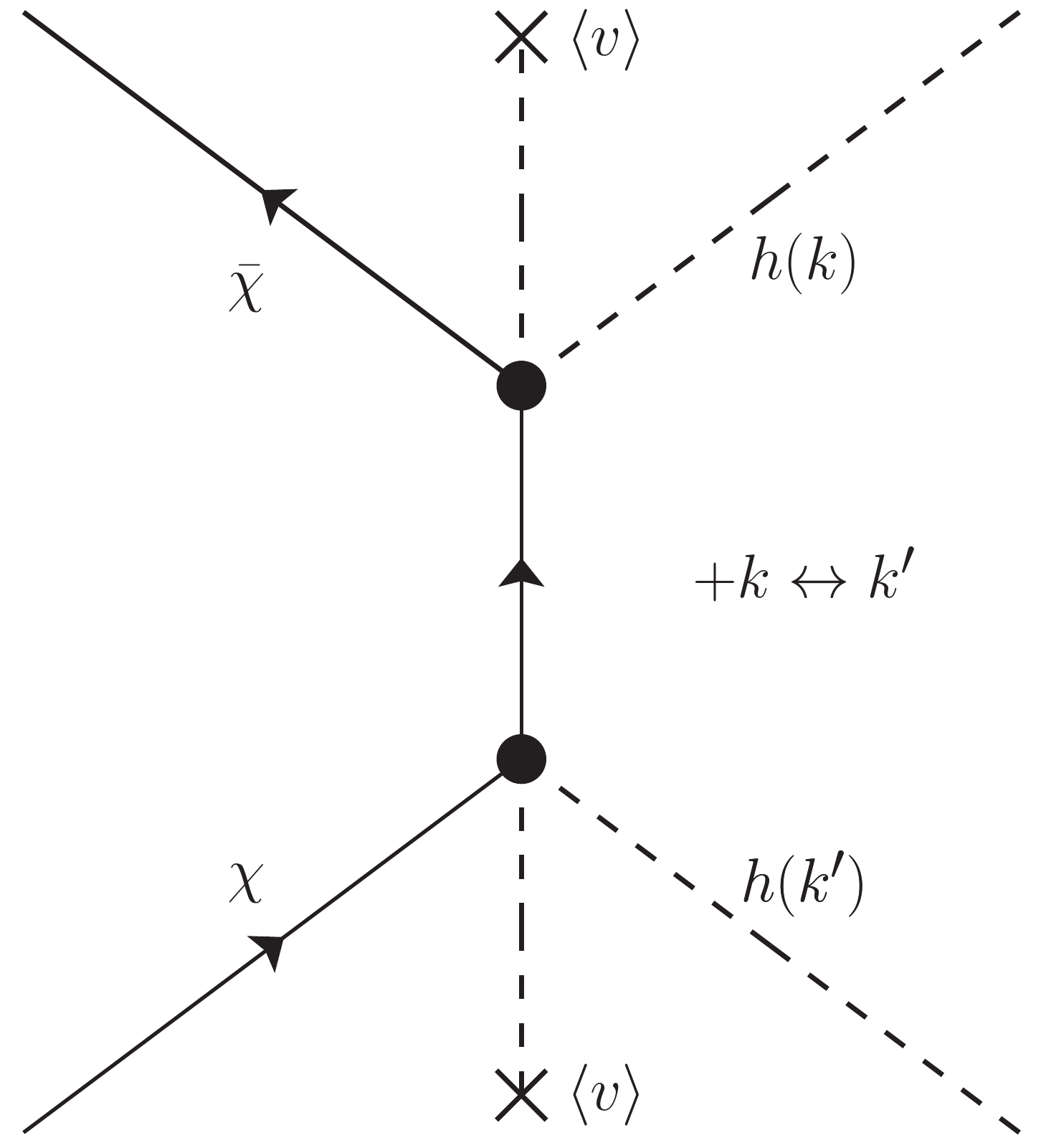}
    \caption{Next-order diagrams [$\mathcal{O}(\Lambda^{-2})$] for the process $\bar{\chi}\chi\rightarrow hh$ resulting from the Higgs-portal operator $H^\dagger H$. \label{fig:TANDU}}
\end{minipage}
\end{figure}

Although the annihilation cross sections for the various channels depend on $\Lambda$ and $\xi$, the branching fractions only depend on $s$ (equal to $4M^2$ in the NR limit);\footnote{This is an exact statement only in the NR limit. The $\Gamma_h$-dependent term in the numerator of $\Sigma_{hh}$ in \eqref{eq:s1} does not cancel in the ratio when we compute the BR, and generally depends on both $\Lambda$ and $\xi$. However, the $hh$ channel is only open in the NR limit for $M \geq m_h$, where the width is independent of any exotic contribution. More generally, there is a dependence on $\Lambda, \xi$ in the BR for $M<m_h/2$, but only when the cross-sections are considered at $s>4m_h^2$ which suppresses this dependence by at least $(\Gamma_h/m_h)^2$.}  these are shown as a function of $M$ in the NR limit in figure \ref{fig:BR}.  Above $W^+W^-$ threshold, the largest branching fraction is to $W^+W^-$, with the branching fractions to $ZZ$ and $hh$, where kinematically allowed, smaller by a factor of a few.\footnote{In the limit of large $M$, the ratios are $W^+W^-:ZZ:hh = 2:1:1$, as one would expect by the Goldstone Boson Equivalence Theorem.}  Below $W^+W^-$ threshold the only annihilation channel is to fermion pairs, predominately to the largest mass kinematically allowed.

Note that the cross sections depend on $\cos^2\xi$ and $\sin^2\xi$.  However this will no longer be true at higher order in $\Lambda^{-1}$. For instance, at higher order in $\Lambda^{-1}$ for the $hh$ final-state there are the additional tree-level $t$- and $u$-channel diagrams illustrated in figure \ref{fig:TANDU}. If we include these diagrams, $\Sigma_{hh}(s;M,m_h)$ becomes significantly more complicated:\footnote{This result was derived with the aid of \textsc{FeynCalc} \cite{FeynCalc} and verified numerically at some sample parameter values using the \textsc{FeynRules} \cite{Alloul:2013bka} and \textsc{CalcHEP} \cite{Belyaev:2012qa} pipeline.}
\begin{align}
& \Sigma_{hh}(s;M,m_h) \nonumber \\& = 
  \frac{1}{4} \frac{s}{\Lambda^2} \frac{\bigg(1+2 m_h^2/s\bigg)^2 + \bigg(\Gamma_h m_h/s \bigg)^2}{\bigg(1-m_h^2/s\bigg)^2+ \bigg( \Gamma_h m_h/s \bigg)^2}  \left[ \cos^2\xi \left( 1 - \frac{4M^2}{s} \right) + \sin^2\xi \right] \nonumber \\[2ex]
  &+ \frac{ 2 M \vev^2 \cos\xi}{\Lambda^3} \frac{  \bigg(1-m_h^2/s\bigg) \bigg(1+2m_h^2/s\bigg) +  \bigg(\Gamma_h m_h/s\bigg)^2  }{ \bigg(1-m_h^2/s\bigg)^2 +  \bigg(\Gamma_h m_h/s\bigg)^2} \nonumber \\ &\qquad \qquad \times \left[ 1 + \frac{1}{\beta}  \left(  1 - \frac{8 M^2}{s} \cos^2\xi + \frac{2 m_h^2}{s} \right) \tanh ^{-1}\left(\frac{\beta}{1-2 m_h^2/s}\right) \right] \nonumber \\[2ex]
   & -\frac{\vev^4}{2\Lambda^4} \left[ \frac{M^2}{s} \left(1-\frac{4 m_h^2}{s}\right)+\frac{m_h^4}{s^2} \right]^{-1}  \left[ \frac{16M^4}{s^2} \cos^4 \xi  + \frac{2 M^2}{s} \left(1-\frac{4m_h^2}{s}\left(1+\cos^2\xi\right)\right)  +\frac{3
   m_h^4}{s^2} \right] \nonumber \\[2ex]
   &+\frac{\vev^4}{\Lambda^4} \beta^{-1}\left(1-\frac{2m_h^2}{s}\right)^{-1} \left[ 1 - \frac{4m_h^2}{s} + \frac{6m_h^4}{s^2} + \frac{16M^2}{s} \left( 1- \frac{m_h^2}{s} \right)\cos^2 \xi - \frac{32M^4}{s^2}\cos^4\xi \right] \nonumber \\ & \qquad \qquad \times \tanh ^{-1}\left(\frac{\beta}{1-2 m_h^2/s}\right) 
\label{eq:SANDU}
\end{align}
where we have defined
\begin{equation}
\beta(s;M,m_h) \equiv \sqrt{\left(1-4M^2/s\right) \left(1-4 m_h^2/s\right) }.
\end{equation}
Note that (contact + $s$) -- $(t,u)$ interference term in \eqref{eq:SANDU} is proportional to $\cos\xi$, while all other terms with $\xi$ dependence are proportional to $\cos^2\xi$ or $\sin^2\xi$. 

This is however just one example of how higher order effects in the EFT suppression scale $\Lambda$ can arise. Within the context of any UV completion, the low-energy EFT will contain a tower of operators beginning at $\Lambda^{-1}$, with other operators in the tower suppressed by higher powers of $E^*/\Lambda$ where $E^*$ is some relevant energy scale (e.g., the momentum of the Higgs, the Higgs vev, etc.). Some of these operators will of course be more important than others for a particular application, but generically, their presence implies that a result (such as \eqref{eq:SANDU}) computed beyond leading order in $\Lambda^{-1}$ using only the couplings arising from the lowest order effective operator is not necessarily complete to that order in $\Lambda^{-1}$, but is merely indicative. 

Absent motivation to the contrary then,\footnote{\label{foot:HOT} A good example of where some higher-order-in-$\Lambda^{-1}$ effects can be much more important than expected and thus should be retained, is furnished by the chiral rotation discussion above: for initially pure pseudoscalar coupling, the coefficient of the scalar coupling which is generated upon rotation is one power higher in $\Lambda^{-1}$ than the leading order coefficient of the pseudoscalar coupling. However, simply neglecting this effect can be a grave error when considering direct detection, where velocity suppression applies to the pseudoscalar coupling only. In order to neglect the higher-order-in-$\Lambda^{-1}$ coefficient of the scalar coupling, it would have to be much smaller than $v^2 \sim 10^{-6}$. For the parameter space we consider though, this never occurs.} we should really only keep the lowest order term, using the known, if incomplete, higher order terms as a way to gauge whether we trust the EFT in any given region of parameter space: if the numerical coefficients for the higher order terms are ``too large'' we should be wary of trusting the lowest-order approximation and must be alert to the possibility that neglected contributions may actually be important.  

In this spirit, at fixed $(M,\xi)$, we will mostly work to lowest order in $\Lambda^{-1}$, but occasionally we will present results using all the terms in \eqref{eq:SANDU} to illustrate the potential magnitude of higher-order terms in $\Lambda^{-1}$.

\section{The Width of the Higgs \label{sec:width}}

For $m_\chi<m_h/2$, the width of the Higgs will differ from the SM value because it is necessary to include the process $h\rightarrow \bar{\chi}\chi$. The presence of this exotic or `invisible' contribution to the Higgs width implies a nontrivial constant on light DM (see e.g.\ refs.\ \cite{Belanger:2013xza,Espinosa:2012vu,Greljo:2013wja,Englert:2011aa,deSimone:2014pda,
Kanemura:2010sh}).  A simple tree-level computation of the partial decay width of the Higgs to a $\bar{\chi}\chi$ pair yields the result (taking $m_h = 126$ GeV)
\begin{align}
\Gamma_{h\rightarrow \bar{\chi}\chi} &= \frac{m_h}{8\pi} \frac{\vev^2}{\Lambda^2} \sqrt{ 1 - \frac{ 4M^2}{m_h^2} } \left[ 1 - \frac{4M^2}{m_h^2} \cos^2\xi \right] \nonumber \\
&= \left( 3.034\times 10^2\ \text{MeV} \right) \times \left( \frac{\text{1 TeV}}{\Lambda} \right)^2 \sqrt{ 1 - \frac{ 4M^2}{m_h^2} } \left[ 1 - \frac{4M^2}{m_h^2} \cos^2\xi \right].
\label{eq:gamma_h}
\end{align} 
Given that a recent CMS result \cite{CMS-PAS-HIG-14-002} which utilizes far off-shell Higgs decaying via $ZZ$ to 4 leptons or 2 leptons and 2 neutrinos limits the total Higgs width to $\Gamma_{h,\text{ total}}^{95\% \text{ CL UL}} \leq 17.4$ MeV at 95\% confidence, we expect a strong constraint here. One can actually do even better than this limit. The invisible branching fraction for Higgs induced by this decay is defined by
\begin{align}
\mathcal{B}_{\text{inv}} = \frac{ \Gamma_{h\rightarrow \bar{\chi}\chi} }{ \Gamma_{\text{SM}} + \Gamma_{h\rightarrow \bar{\chi}\chi} },
\end{align}
where the theoretical value of the total width of a 126 GeV Higgs boson is $\Gamma_{\text{SM}} = 4.21$ MeV \cite{Heinemeyer:2013tqa}. This is an extremely small value compared to the fiducial partial width to $\bar{\chi}\chi$ shown in \eqref{eq:gamma_h}, which implies that even for fairly modest limits on $\mathcal{B}_{\text{inv}}$ the resulting constraints will be very strong in the kinematically allowed region. 

The present best limits on the invisible branching ratio come from a global fit to Higgs data and are $ \mathcal{B}_{\text{inv}} < 0.19(0.38)$ \cite{Belanger:2013xza} for the case where the Higgs couplings are fixed to their theoretical SM values (allowed to float freely in a global fit); these results were computed using $m_h = 125.5$ GeV, but should be very similar for our choice of $m_h = 126$ GeV. Although we do not utilize them further in our analysis, we also note that searches for invisible Higgs decay in the ZH associated production mode at ATLAS \cite{Aad:2014iia} and in the ZH associated production and vector boson fusion (VBF) modes at CMS \cite{Chatrchyan:2014tja} yield 95\% confidence level upper limits on ${\cal B}_{\text{inv}}$ of 0.75 and 0.58, respectively, assuming in both cases SM production cross-sections and Higgs masses of 125.5 GeV and 125 GeV, respectively.

The invisible width is halved for the Majorana case: while the conventional factor of $1/2$ in the Lagrangian ensures the same amplitude for decay as for the Dirac case, the Majorana fermions are now both in the final state necessitating an additional factor of $1/2$ to avoid double counting the phase-space.

Combined with the values of $\Lambda$ required for the correct relic abundance (see below), the resulting invisible-width limits from the global fit analysis from ref.~\cite{Belanger:2013xza} on the allowed values of $M$ are approximately independent of the value of $\xi$ and are $M \gtrsim 56.8(56.2)$ GeV for the Dirac case and $M \gtrsim 55.3(54.6)$ GeV for the Majorana case for fixed (floating) couplings. The constraints from the CMS limit on the total width \cite{CMS-PAS-HIG-14-002} are only slightly weaker and also approximately independent of $\xi$, limiting the DM mass to be $M \gtrsim 55.7$ GeV for the Dirac case and $M \gtrsim 53.8$ GeV for the Majorana case.

\section{The Relic Abundance \label{sec:relic}}
The familiar Boltzmann equation \cite{kolb1994early,Gondolo:1990dk,Srednicki:1988ce} for a single species of number density $n$ (the particle density only; \emph{not} the combined particle and anti-particle density) whose equilibrium abundance is $n_\mathrm{EQ}$, undergoing only annihilations with itself or its anti-particle is written as:\footnote{The same form obtains for both particle-particle (e.g., Majorana fermion), and particle-anti-particle (e.g., Dirac fermion) annihilations because while there is a factor of 2 on the RHS to account for the loss of two particles per annihilation in the former case, it cancels a factor of $1/2$ to avoid double-counting the \emph{initial} state phase space --- that is, the factor of $1/2$ which accounts for the combinatoric factor of $N(N-1)/2 \approx N^2/2$ possible pairs of interactions given $N$ total particles undergoing annihilation among themselves \cite{Srednicki:1988ce}.}
\begin{align}
\dot{n} + 3Hn = - \langle \sigma v_{\text{M\o ller}} \rangle\left[ n^2 - n_\mathrm{EQ}^2 \right]
\end{align}
where $\langle \sigma v_{\text{M\o ller}} \rangle$ is the thermal average of $ \sigma v_{\text{M\o ller}} (s)$, given by \cite{Gondolo:1990dk}
\begin{align}
\langle \sigma v_{\text{M\o ller}} \rangle = \left[ 8M^4 T K_2^2(M/T) \right]^{-1} \int_{4M^2}^{\infty} \sigma(s) \ ( s - 4M^2 ) \ \sqrt{s}\ K_1(\sqrt{s}/T) \ ds,
\end{align}
and $K_{1,2}$ are modified Bessel functions (this expression assumes Boltzmann statistics for the DM at freeze-out).\footnote{We note that for the large $M/T$ regime, the formula as shown can be numerically problematic as it involves the ratio of two exponentially small numbers and a large-argument asymptotic expansion of the Bessel functions is necessary.} With the usual definition $Y\equiv n/s$ ($s$ the entropy density), and $Y_\infty$ denoting the post-freeze-out value of $Y$, the present ratio of the WIMP mass density to the present critical density $\rho_c=3H_0^2/8\pi G$ is \cite{kolb1994early}
\begin{equation}
\Omega = N_{DM} \frac{M s_0}{\rho_c} Y_\infty \ .
\end{equation}
The present value of the entropy density is $s_0 = 2891$ cm$^{-3}$ \cite{Beringer:1900zz} and $N_{DM} = 1\ (2)$ for (non-)self-conjugate DM.\footnote{This is the only place where the difference between Majorana and Dirac fermions enters in this computation \cite{Srednicki:1988ce}.}  Observationally, the DM relic abundance is determined to be $\Omega h^2 = 0.1186(31)$ \cite{Ade:2013zuv}, where $H_0=100h$ km s$^{-1}$ Mpc$^{-1}$.

\begin{figure}[t]
\begin{center}
\includegraphics[width=.475\linewidth,keepaspectratio]{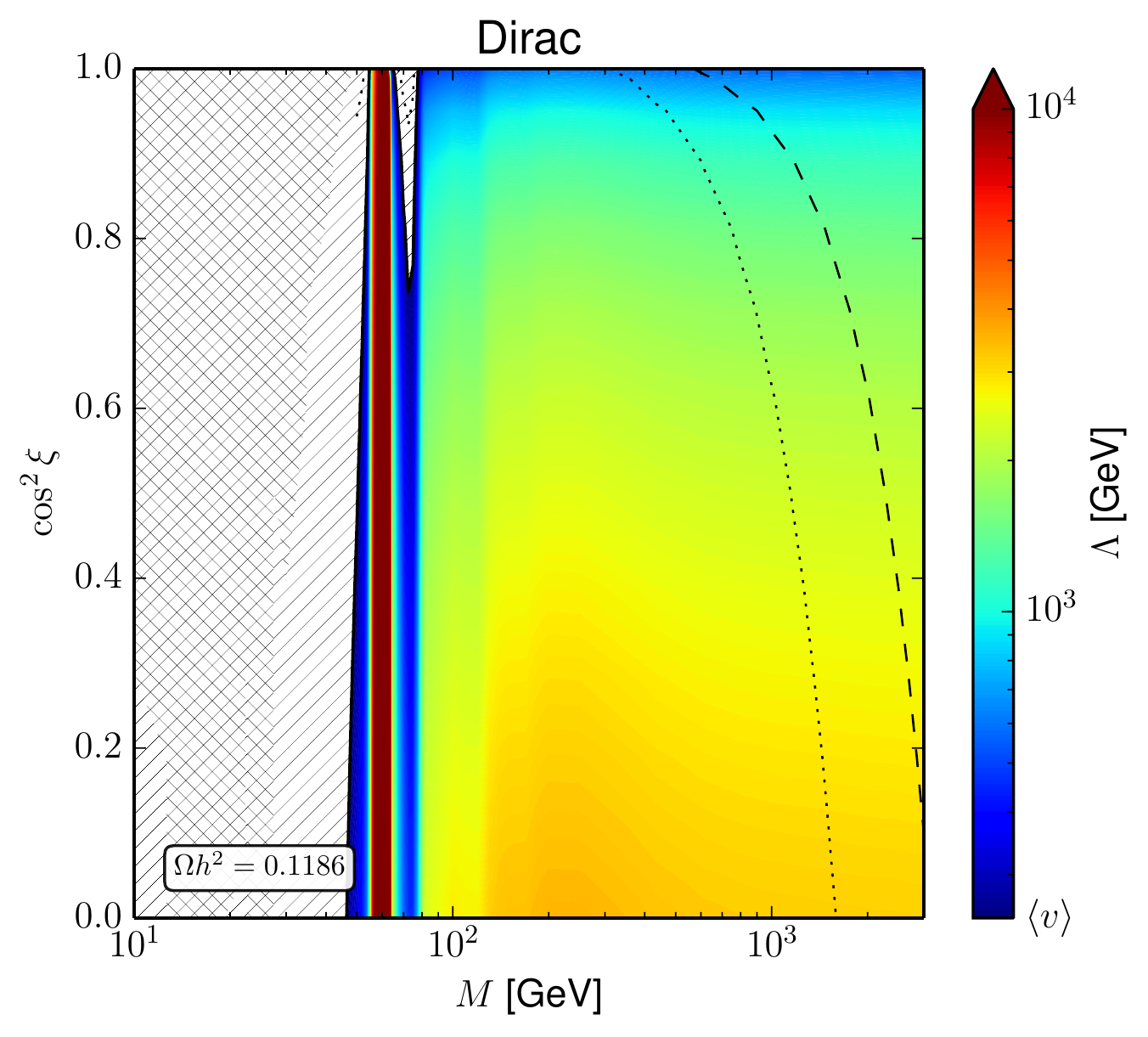} 
\includegraphics[width=.475\linewidth,keepaspectratio]{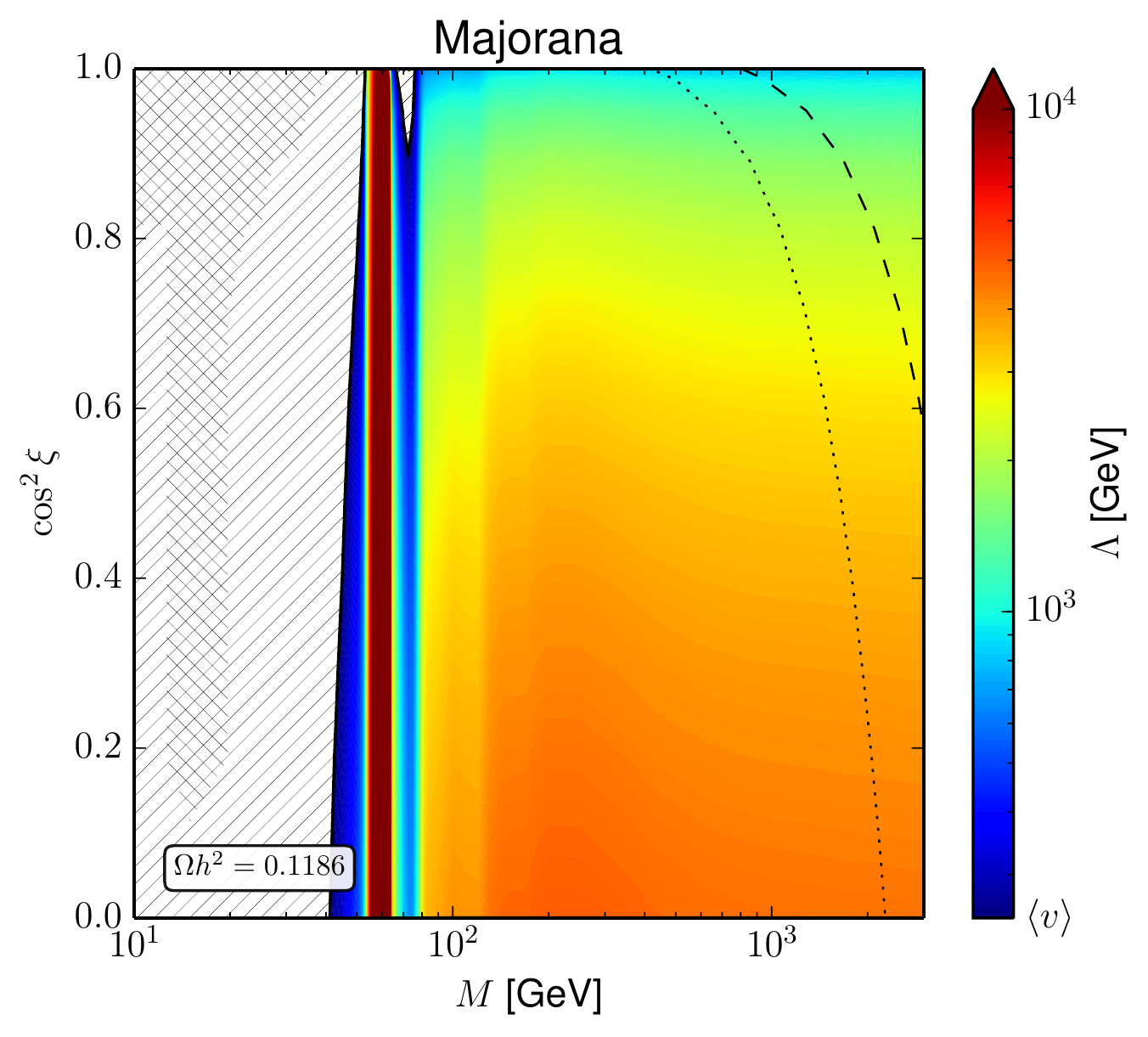}\\
\caption{ \label{fig:DM}  These colormaps give the (interpolated) values of the EFT suppression scale $\Lambda$ required for the correct relic abundance, $\Omega h^2 = 0.1186$, for various values of $(\cos^2\xi,M)$. Apart from a region near $M\sim m_h/2$ where the presence of the $s$-channel resonance forces the value of $\Lambda$ to exceed 10 TeV, and the troughs on either side of this resonance, the values of $\Lambda$ are generically between a few hundred GeV and a few TeV, with smaller values required for more scalar cases ($\cos^2\xi \rightarrow 1$) due to the velocity suppression ($v^2\sim0.3$ at freeze-out \cite{kolb1994early}) of the pure-scalar annihilation channel cross-section. Note that there may be regions where $\Lambda$ is too small for the EFT to be taken seriously: in the singly hatched region, $\Lambda\lesssim \vev$ (see discussion in the text), and the dotted and dashed lines indicate, respectively, where $\Lambda\lesssim 2M$ and $\Lambda \lesssim M$ (for reference, the simplest perturbatively unitary UV completion requires the scale of new physics to be $\geq M/2\pi$ \cite{Busoni:2014ty}; although in our parametrization $\Lambda$ is not necessarily exactly this scale, it is of the same order of magnitude). The doubly hashed region is where no $\Lambda$ can be found which gives the correct relic density for the chosen $M$ and $\cos^2\xi$ (the boundaries of this region as shown are not entirely smooth due to sampling effects on the computation grid and should thus be taken as indicative only; also, as they always lie in the regions where $\Lambda < \vev$, their validity is in any event open to question).}
\end{center}
\end{figure}

There are three parameters in the EFT: $\Lambda$, $M$, and $\xi$.  If we consider only the lowest non-vanishing order in $\Lambda^{-1}$, the cross section depends only upon $\cos^2\xi$.  The values of $\Lambda$ required to give $\Omega h^2=0.1186$ \cite{Ade:2013zuv} are shown as a function of $M$ and $\cos^2\xi$ in figure \ref{fig:DM}.  In agreement with previous literature (see e.g., ref.\ \cite{LopezHonorez:2012kv}), we find that for $M<m_h/2$ if there is at least one allowed value of $\Lambda$ giving the correct relic density, then there are typically two such allowed values\footnote{Note that this is an independent concern from that mentioned above in the context of discussing the mapping between $(M_0,\theta)$ and $(M,\xi)$.} for $\Lambda$ since the annihilation cross-sections scale parametrically as $ 1 / \sigma \sim \Lambda^{2}  \left[ A + B \Lambda^{-4} \right] \sim \left[ A \Lambda^2 + B \Lambda^{-2} \right]$ for some $A,B$, due to the presence of the exotic contribution to the Higgs width which enters in the denominator of the $s$-channel resonance peak in \eqref{eq:s1}. However, one solution is typically a factor of a few or more smaller than the other, and we always take the larger value of $\Lambda$. This is done for two reasons: a) the smaller value of $\Lambda$ can run far below $\vev$, which is a region where we do not particularly trust the lowest-order EFT results due to i) possible large corrections proportional to $(\vev/\Lambda)^n$ from neglected higher order operators (see discussion below \eqref{eq:SANDU} --- although the corrections from \eqref{eq:SANDU} are not themselves relevant in this mass range), ii) possible perturbative unitarity issues, and iii) possible issues with having implicitly integrated out physics at or below the weak scale (although, see footnote \ref{ft:scale_caveat}), while keeping other weak-scale physics in the theory explicitly; and b) the largest value of $\Lambda$ implies the loosest constraints on any given operator from present-day experimental data and this gives the most conservative approach to setting exclusion bounds. 

\begin{figure}[t]
\begin{center}
\includegraphics[width=.475\linewidth,keepaspectratio]{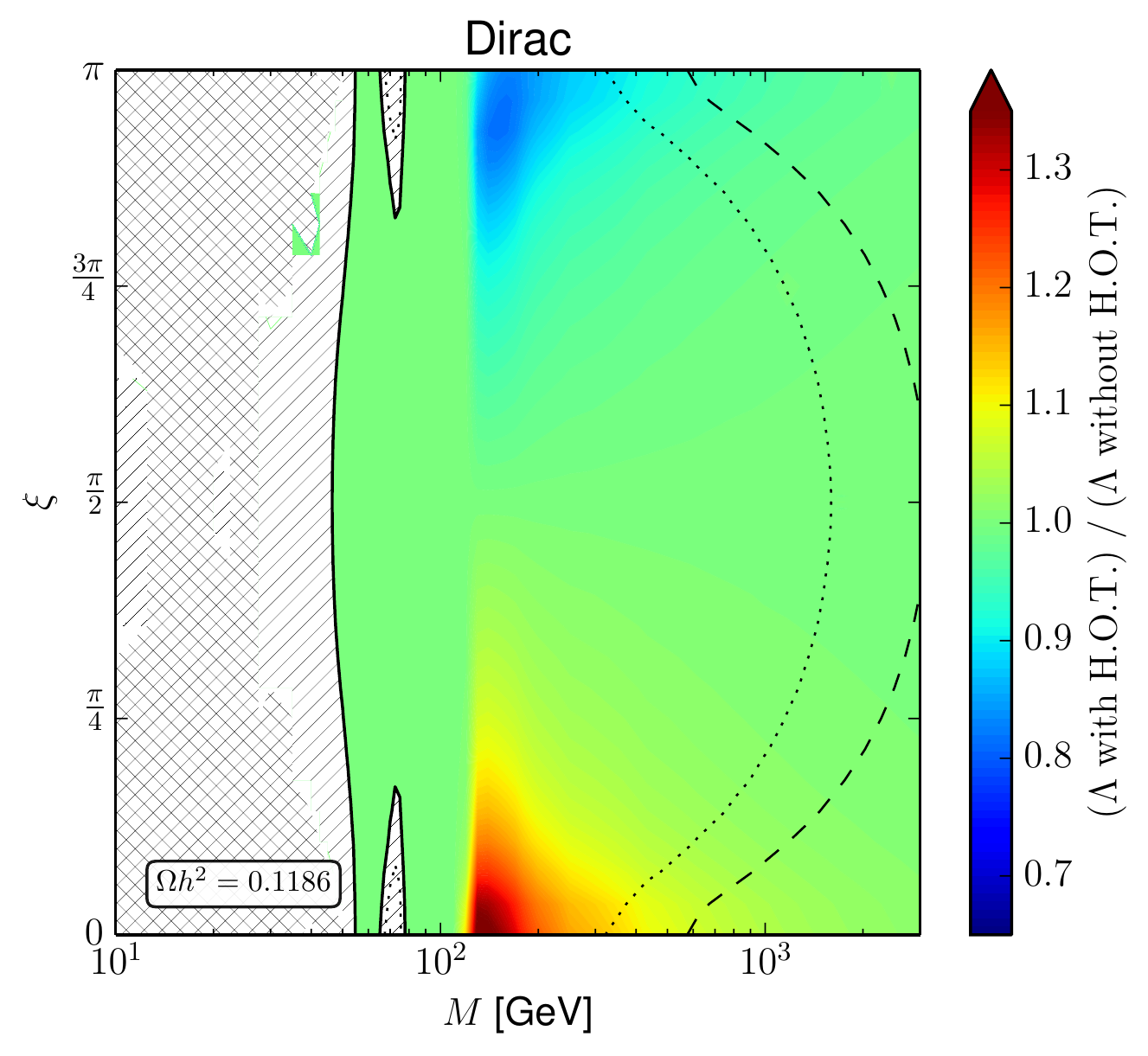} 
\end{center}
\caption{ \label{fig:HOT}  The ratio of the values of the EFT suppression scale $\Lambda$ required for the correct relic abundance for various values of $(M,\xi)$ assuming that $\Omega h^2 = 0.1186$ for the two cases of 1) when the higher-order terms (H.O.T.) in the cross section of \eqref{eq:SANDU} are included and 2) when they are neglected. The differences are fairly small over most of the parameter space, and even in regions where they differ by $\mathcal{O}(30\%)$, we have explicitly checked that their ultimate impact on the exclusion bounds we set (see below) is very small. See the caption of figure \ref{fig:DM} for description of the hashed regions and dotted/dashed lines. As this figure is mainly for illustrative purposes, we do not present independent results for the Majorana case.}
\end{figure}

There are also regions where for a given $M$ and $\cos^2\xi$, there is no value of $\Lambda$ that will give the correct relic abundance (the cross-section has an upper bound as a function of $\Lambda$ with all other parameters held fixed). This is illustrated by the double-hash regions of figure \ref{fig:DM}.   The single-hashed regions denote where the values of $\Lambda$ to give the correct $\Omega h^2$ are smaller than $\vev$; for the variety of reasons already advanced above, we expect that our lowest-order EFT results are not entirely trustworthy in this region, and we thus conservatively choose not to present results here.

In figure \ref{fig:DM} one clearly sees the region where resonant annihilation occurs around $M \simeq m_h/2$.  In this region a very large value of $\Lambda\gtrsim 10$ TeV is required. This is the ``resonant Higgs Portal'' scenario of ref.\ \cite{LopezHonorez:2012kv} (see also ref.\ \cite{Ibe:2008ye}).

So far we have ignored possible higher-order terms in $\Lambda^{-1}$ in presenting our results.  As we have seen from one of the possible contributions considered in \eqref{eq:SANDU}, higher-order terms may depend on $\cos\xi$ and not simply $\cos^2\xi$.  An illustration of the possible magnitude of these terms, we have calculated the values of $\Lambda$ necessary to arrive at the correct relic density for various values of $\xi$ and $M$ including the higher-order terms in \eqref{eq:SANDU}.  The result is illustrated in figure \ref{fig:HOT}; the effect of higher-order terms is small, and we henceforth will ignore them.

\section{Direct Detection \label{sec:direct}}

Direct detection constraints are particularly important to consider for these Higgs portal operators \cite{Djouadi:2011aa,Greljo:2013wja,LopezHonorez:2012kv,deSimone:2014pda,Kanemura:2010sh,
Kim:2006af}. The relevant process is the $h$-mediated $t$-channel elastic scattering of WIMPs on nucleons. Using the interaction Lagrangian of \eqref{eq:ta} together with the Higgs-quark coupling term from the SM Lagrangian yields
\begin{equation}
\Lag \supset - \sum_q \frac{m_q}{\vev} \ h \ \bar{q} q + \Lambda^{-1} \left[ \cos\xi\ \Sc + \sin\xi \ \PSc \right] \vev \ h .
\end{equation}
Since the momentum transfer in the scattering process is typically less than an MeV, very much less than the Higgs mass (126 GeV), the Higgs can be integrated out to obtain the effective operator connecting DM to quarks,
\begin{equation}
\Lag_{\text{eff}}^{\text{direct detection} } \supset - \sum_q \frac{1}{m_h^2} \ \frac{m_q}{\Lambda} \ \bar{q} q \ \left[ \cos\xi\ \Sc + \sin\xi \ \PSc \right],
\end{equation}
where the factor of $m_h^{-2}$ comes from integrating out the $h$ propagator.  The sum runs over all quarks, with the heavier quarks entering the direct detection process through triangle diagrams which induce effective couplings of the $h$ to the gluons in the nucleon; this is usually accounted for by writing an effective matrix element for the heavy quarks in the nucleon (see e.g.\ ref.~\cite{Agrawal2010} and references therein).

Following the well-known procedure to extract the nuclear matrix elements (see e.g.\ appendix B of ref.\ \cite{Agrawal2010}), the result for the spin-averaged and phase-space integrated $S$-matrix element is\footnote{Strictly speaking, the part of the cross-section appearing as velocity suppressed here is momentum-transfer-suppressed; only in the elastic scattering case is this equivalent to velocity suppression.}
\begin{equation}
\left\langle | \mathcal{M} | \right\rangle \equiv \int \frac{ d\Omega }{ 4\pi } \frac{1}{4} \sum_{\text{spins}} \left| \mathcal{M} \right|^2 = 16 \left(M + M_N\right)^2 \left( \frac{\mu_{ \chi N} }{ m_h^2 } \right)^2 \left( \frac{ f_N }{ \Lambda } \right)^2 \left[ \cos^2\xi +  \frac{1}{2} \left(\frac{\mu_{\chi N}}{M} \right)^2  \nu_\chi^2 \right],
\end{equation}
where
$f_N \equiv M_N \left( \sum_{q=u,d,s} f_{Tq}^{(N)} + \frac{2}{9} f_{TG}^{(N)} \right) \approx 0.35 M_N \approx 0.33 \text{GeV}$
 \cite{Ellis:2000ds,Agrawal2010} is the nuclear matrix element accounting for the quark (and gluon, through heavy quark triangle diagrams) content of the nucleon to which the Higgs couples, $\mu_{\chi N} = M M_N / ( M_N + M )$ is the reduced mass of the WIMP-nucleon system,  and $\nu_\chi$ is the DM speed in the nucleon rest frame (the mass ratio in front of the squared velocity arising when one goes from the CoM frame to the nucleon rest frame) (see also ref.~\cite{Tsai:2013bt}). We will take the DM speed to be $\nu_\chi \sim 220$ km/s in the earth rest-frame; a proper treatment would require an averaging over the DM velocity distribution already in the extraction of the cross-section exclusion bound from LUX data, and not \emph{a posteriori} once a bound is extracted, as there are additional velocity-dependent factors which enter the conversion from the differential recoil rate in the detector to a cross-section bound (see e.g. ref.\ \cite{Cerdeno:2010zz}).

The total cross-section is 
\begin{eqnarray}
\sigma_{\text{SI}}^{\chi N} & = & \frac{\left\langle | \mathcal{M} | \right\rangle}{ 16 \pi ( M + M_N )^2 } = \frac{1}{\pi} \left( \frac{\mu_{ \chi N} }{ m_h^2 } \right)^2 \left( \frac{ f_N }{ \Lambda } \right)^2 \left[ \cos^2\xi + \frac{1}{2}\left(\frac{\mu_{\chi N}}{M} \right)^2 \nu_\chi^2 \right] \label{eq:SIsigma_full} \\
& = & 4.7\times10^{-38} \text{cm}^2 \ \left(\frac{M}{\Lambda}\right)^2 \left( \frac{1 \text{ GeV}} {0.94\text{ GeV}+ M} \right)^2 \ \left[ \cos^2\xi + \frac{1}{2} \left(\frac{\mu_{\chi N}}{M} \right)^2 \nu_\chi^2 \right]\ . \label{eq:SIsigma}
\end{eqnarray}

\begin{figure}[t]
\begin{center}
\includegraphics[width=0.45\textwidth]{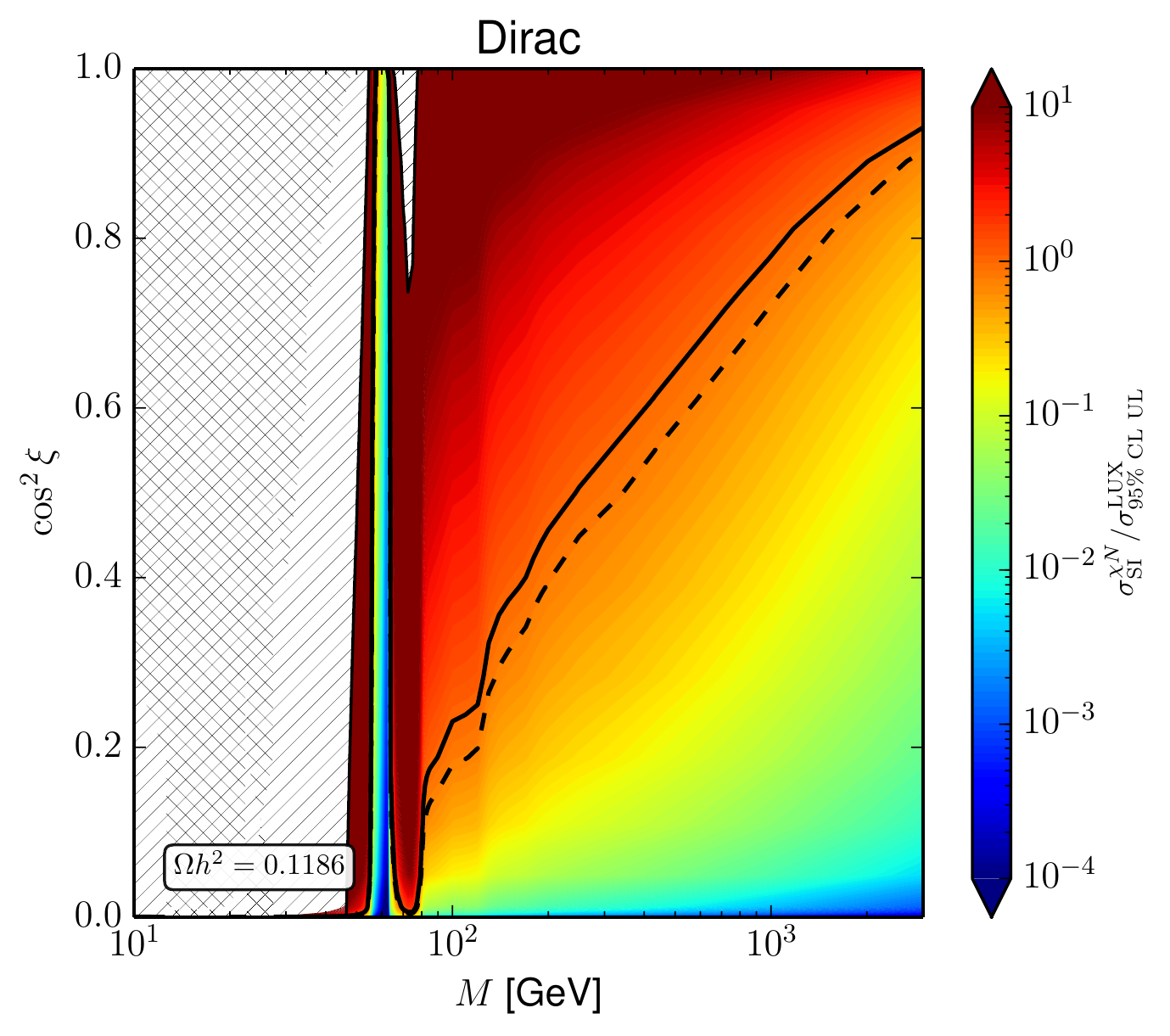}
\includegraphics[width=0.45\textwidth]{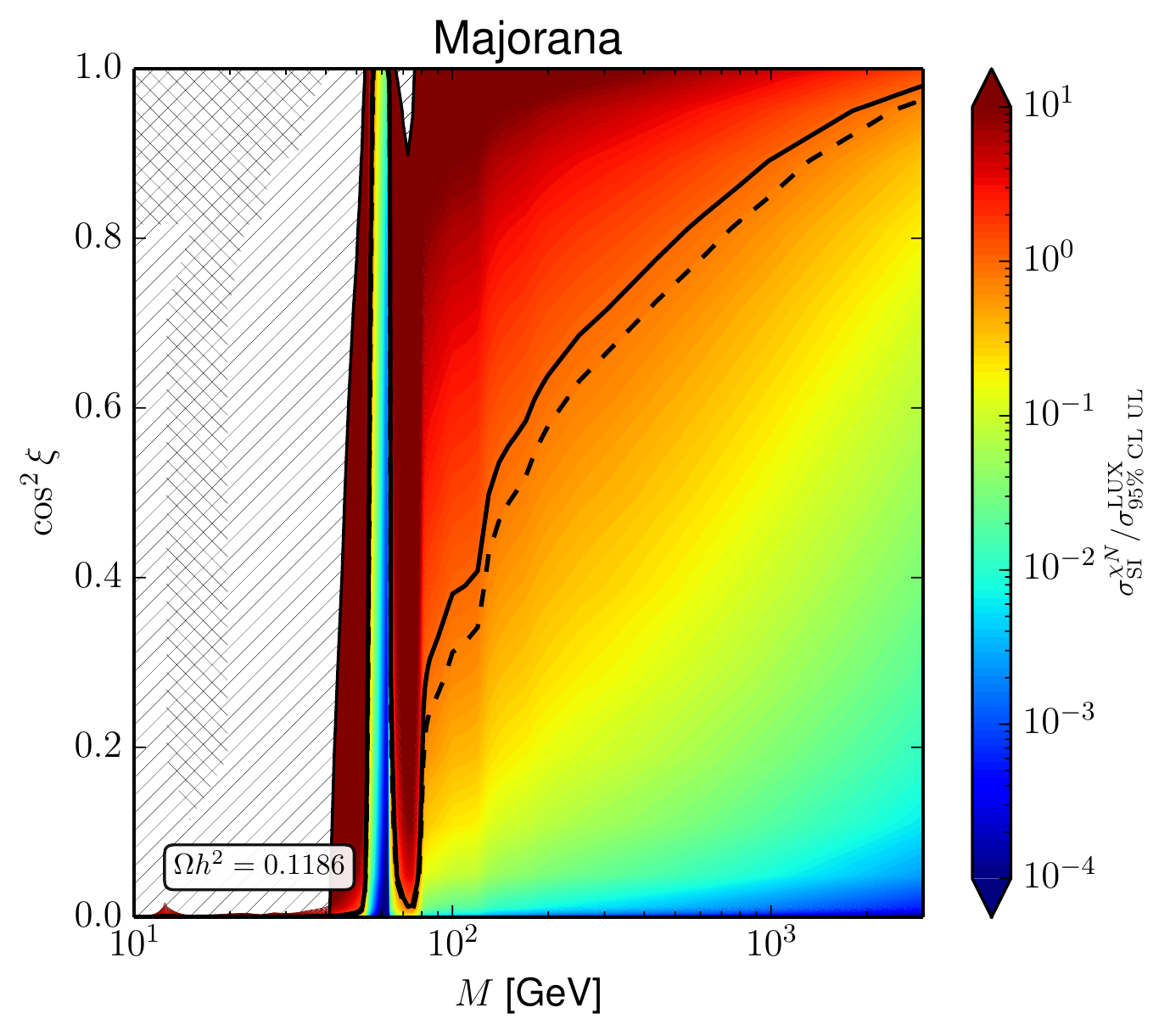}
\end{center}
\caption{ \label{fig:direct_exclusions}    These colormaps represent interpolated values of $\sigma_{\text{SI}}^{\chi N} / \sigma_{\text{95\% CL UL}}^{\text{LUX}}$, with the solid black line showing the equality of the computed cross section and the LUX limit \cite{Akerib:2013tjd} (note that 95\% CL UL from LUX are only available up to 2 TeV from DMTools \cite{DMTools}; we have extrapolated the limit linearly up to 3 TeV --- this is justified since the limit $\sim 1/n_{DM} \sim M$ and since in the data the limit is already scaling approximately linearly in this region). Redder points ``above'' the black line are excluded, bluer points ``below'' the black line are allowed. For reference, the dashed black line is the cognate of the solid black line, except for the 90\% CL UL from LUX: it shows the equality of the computed cross-section and this limit; no other 90\% CL UL contours are shown (90\% CL UL are available up to 3 TeV). Note that the mass region near $M \approx m_h/2$ is allowed for any value of $\xi$: this is the resonant Higgs portal scenario \cite{LopezHonorez:2012kv}. The singly hatched region is where $\Lambda < \vev$. The doubly hatched region at low mass is where no $\Lambda$ value can be found to obtain the correct relic density.}
\end{figure}

We will compare this to the latest LUX upper limits \cite{Akerib:2013tjd} on the spin-independent WIMP-nucleon cross-section as supplied in numerical form by DMTools \cite{DMTools}. Results are shown in figure \ref{fig:direct_exclusions} for both Dirac and Majorana DM.

\section{Combined Limits \label{sec:combined}}
\begin{figure}
\includegraphics[height=0.45\textwidth]{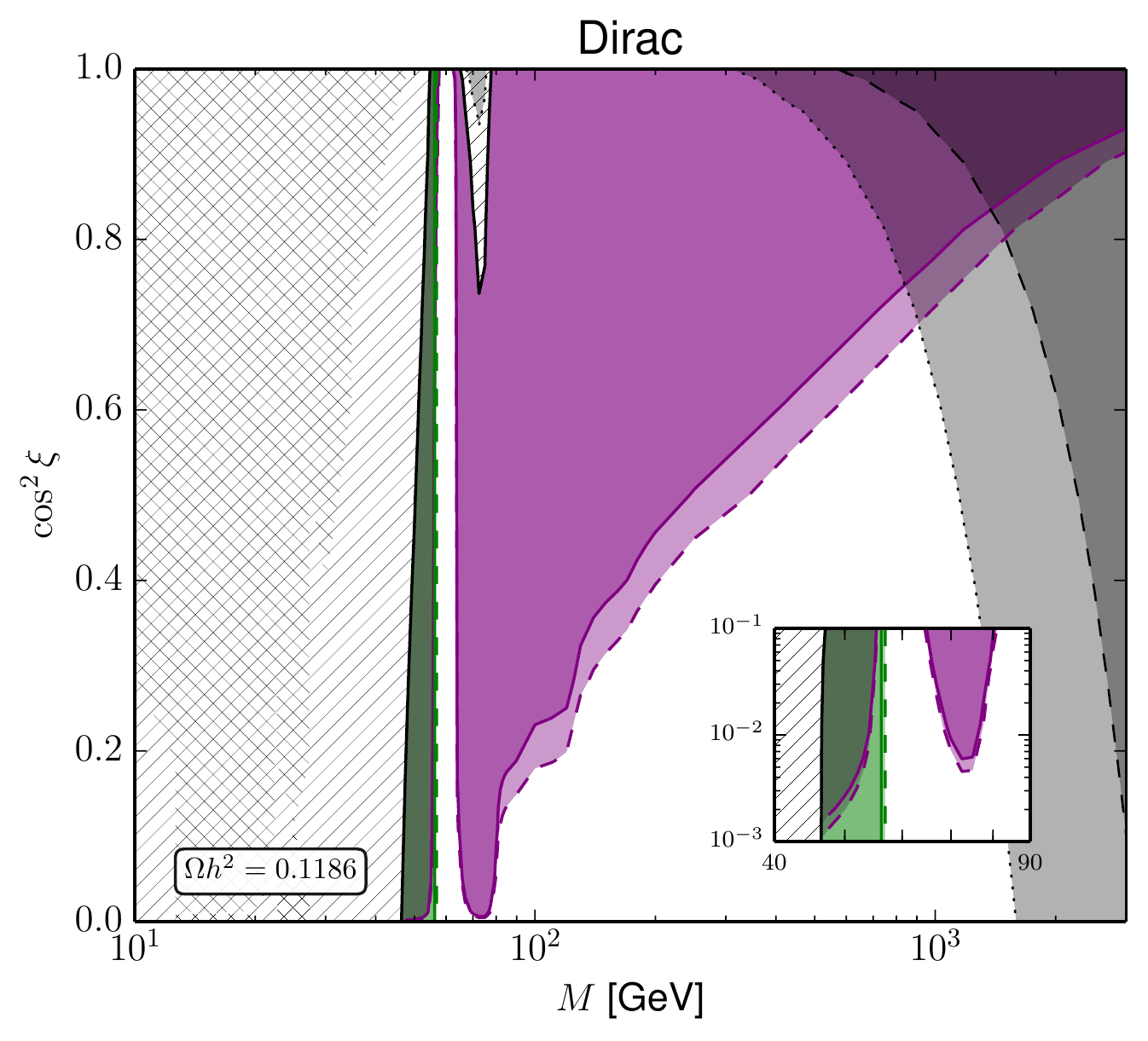}
\includegraphics[height=0.45\textwidth]{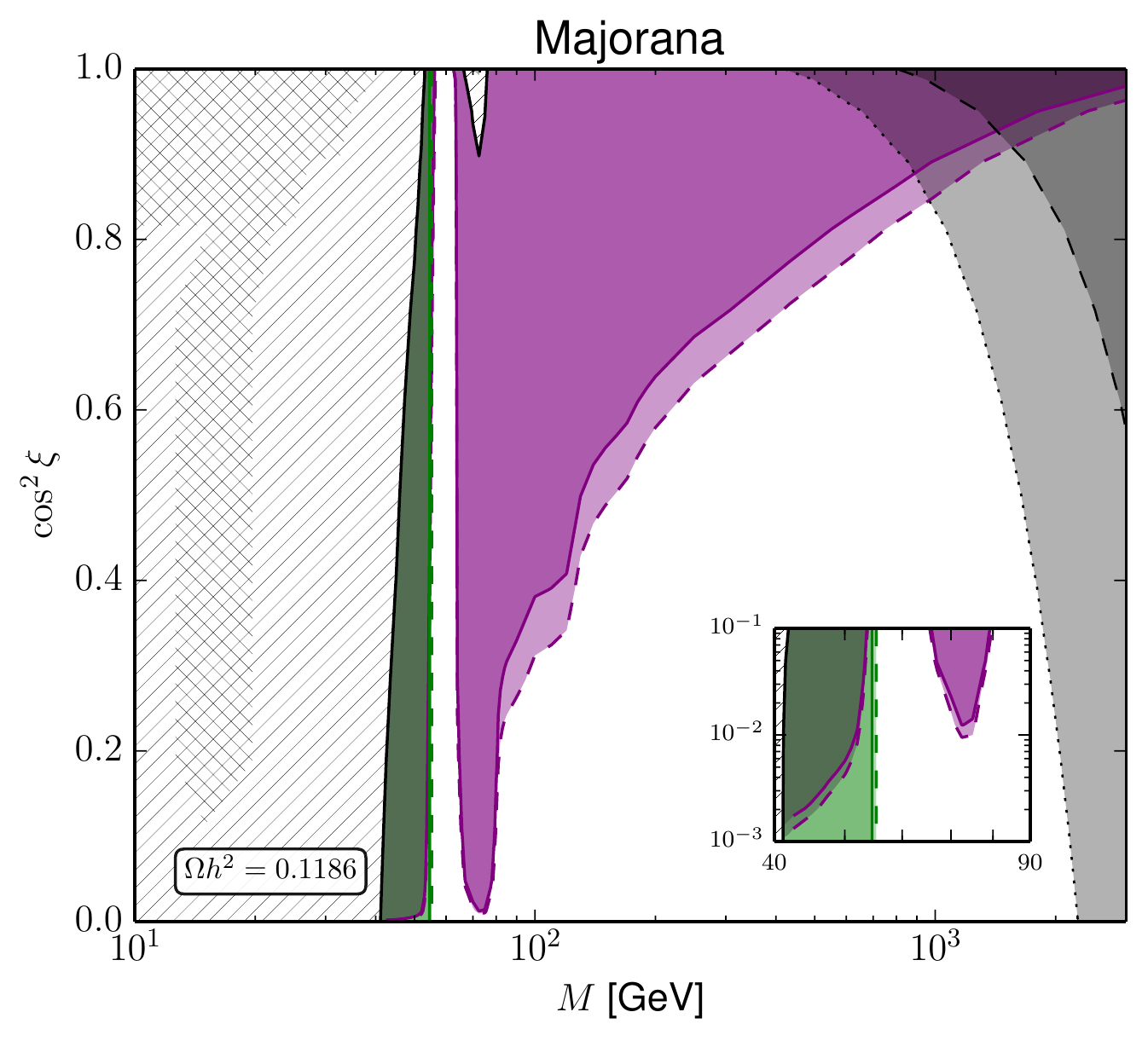}
\caption{ \label{fig:combined_exclusion} Combined limits keeping only $\Lambda^{-2}$ terms in all cross-sections. The grey regions bounded by black lines indicate the regions where $\Lambda \leq 2M$ (dotted line, light grey) and $\Lambda \leq M$ (dashed line, dark grey). The singly hatched region is where $\Lambda < v$. For reference, the simplest perturbative UV completion requires $\Lambda \gtrsim M/2\pi$. The purple shaded regions bounded by purple lines are the regions excluded by LUX \cite{Akerib:2013tjd} direct detections limits at 95\% CL UL (solid line; limit extrapolated between 2 and 3 TeV) and 90\% CL UL (dashed line). The green regions bounded by green lines indicate regions excluded by invisible width constraints arising from a global fit to data \cite{Belanger:2013xza} with the SM-Higgs couplings floating (solid line) or fixed to SM values (dashed line); for clarity, we do not show the limits from the CMS direct width constraints \cite{CMS-PAS-HIG-14-002} as they are slightly weaker. The doubly hatched region at low mass is where no $\Lambda$ value can be found to obtain the correct relic density. The insets show detail for small $\cos^2\xi$ for masses $M \in [40,90]$ GeV and indicate that the coupling must be nearly pure pseudoscalar for masses around 70 GeV to not conflict with data.}
\end{figure}

The combined limits are shown in figure \ref{fig:combined_exclusion} for Dirac and Majorana DM. The inserts are regions where $\cos^2\xi$ is very close to zero and the EFT DM--Higgs coupling is nearly completely pseudoscalar; as discussed above, such a pure pseudoscalar coupling is unnatural. Indirect limits are not included, but are not expected to improve the exclusions shown: since the spectra of particles relevant for indirect detection are fairly featureless continuum spectra, the resulting indirect limits are only constraining for $M\lesssim 30$ GeV once the astrophysical uncertainties are considered \cite{Fedderke:2013pbc}. This mass region is however already strongly excluded by invisible decays of the Higgs. We do however note that in the remaining allowed region, the EFT suppression scale necessary for this scenario to work is in the fairly narrow region $\Lambda \sim $ 1--5 TeV except near the resonance, and this may have interesting implications for collider searches.

\section{Conclusions \label{sec:conclusions}}
In this work, we have examined in an effective field theory approach both Majorana or Dirac SM-singlet fermion dark matter interacting with the SM via some combination of scalar ($\bar{\chi}\chi$) and pseudoscalar ($i\bar{\chi}\gamma_5\chi$) DM operators coupling to the Higgs portal operator $H^\dagger H$. We have performed a systematic scan over DM mass and the ratio of scalar to pseudoscalar coupling strengths, using cosmological measurements of the DM relic density to constraint the EFT suppression scale. We have observed that EWSB necessarily destabilizes a scenario in which the coupling is pure pseudoscalar before EWSB, concluding that this scenario is thus ill-motivated. We have constrained the post-EWSB DM mass and scalar-to-pseudoscalar-coupling ratio with a combination of direct detection bounds from the LUX experiment \cite{Akerib:2013tjd} and with Higgs width constraints as measured by CMS \cite{CMS-PAS-HIG-14-002}, as well as those inferred from a global fit to available Higgs data \cite{Belanger:2013xza}. We find in agreement with ref.~\cite{LopezHonorez:2012kv} that the ``resonant Higgs portal'' scenario in which $M \sim m_h/2$, is still allowed for any admixture of scalar and pseudoscalar couplings (although absent a compelling case from the UV theory for this particular DM mass, this scenario does not seem well motivated). We find that for $M\lesssim 54-56$ GeV, the Higgs portal scenario is ruled out by a combination of direct detection and invisible width constraints independent of the nature of the coupling or Majorana/Dirac nature of the fermionic DM, although in this regime the EFT validity becomes increasingly open to question as $M$ decreases due both to neglected higher order terms and possible perturbative unitarity issues. For masses $m_h/2\lesssim M \lesssim m_W$, the coupling must be almost pure pseudoscalar ($\cos^2\xi \lesssim 1\times 10^{-2}\ (5\times 10^{-3})$ for Dirac (Majorana) cases) to be consistent with present data; this is an ill-motivated scenario due to the accidental relation between parameters $\Lambda M_0 \cos\theta \approx \vev^2/2$ required to achieve it. However, for masses above the threshold for annihilation to $W^+W^-$ (and presumably also for masses slightly below this threshold if we had properly accounted for 3- and 4-body decays through one or two off-shell $W^\pm$) the interaction need not be so finely tuned to be pure pseudoscalar: it suffices for lower $M$ that the coupling is predominantly pseudoscalar, and as the DM mass $M$ is increased, the admixture of scalar coupling allowed increases due to the weakening of the LUX direct detection bounds. Other than in the resonant portal mass region, we find that a pure-scalar Higgs portal coupling is robustly ruled out at at least 95\% confidence for $M$ up to at least 3 TeV for both Majorana and Dirac fermion DM.

While we did not perform an in-depth indirect detection analysis, such limits seem not to hold much promise for strengthening the exclusion bounds on this scenario. Direct searches at colliders in the mass region $M < m_h/2$ are expected to remain weaker than the invisible width limits, while collider signals for larger $M$ cases (specifically, signals with two forward tagging jets and large missing energy (VBF MET), or mono-$X$ and missing energy) may be interesting to examine, but we anticipate that sizable SM backgrounds will make such searches fairly challenging.

\section*{Acknowledgements}
The authors would like to thank Mikhail Solon for useful discussions. This work was supported in part by the Kavli Institute for Cosmological Physics at the University of Chicago through grant NSF PHY-1125897 and an endowment from the Kavli Foundation and its founder Fred Kavli. L.T.W. is supported by the DOE Early Career Award under grant de-sc0003930. The work of E.W.K.\ is supported by the Department of Energy.



\appendix
\section{Selected results in terms of $(M_0,\theta)$ \label{app:selected_results}}

\begin{figure}
\begin{center}
\includegraphics[width = 0.49 \textwidth]{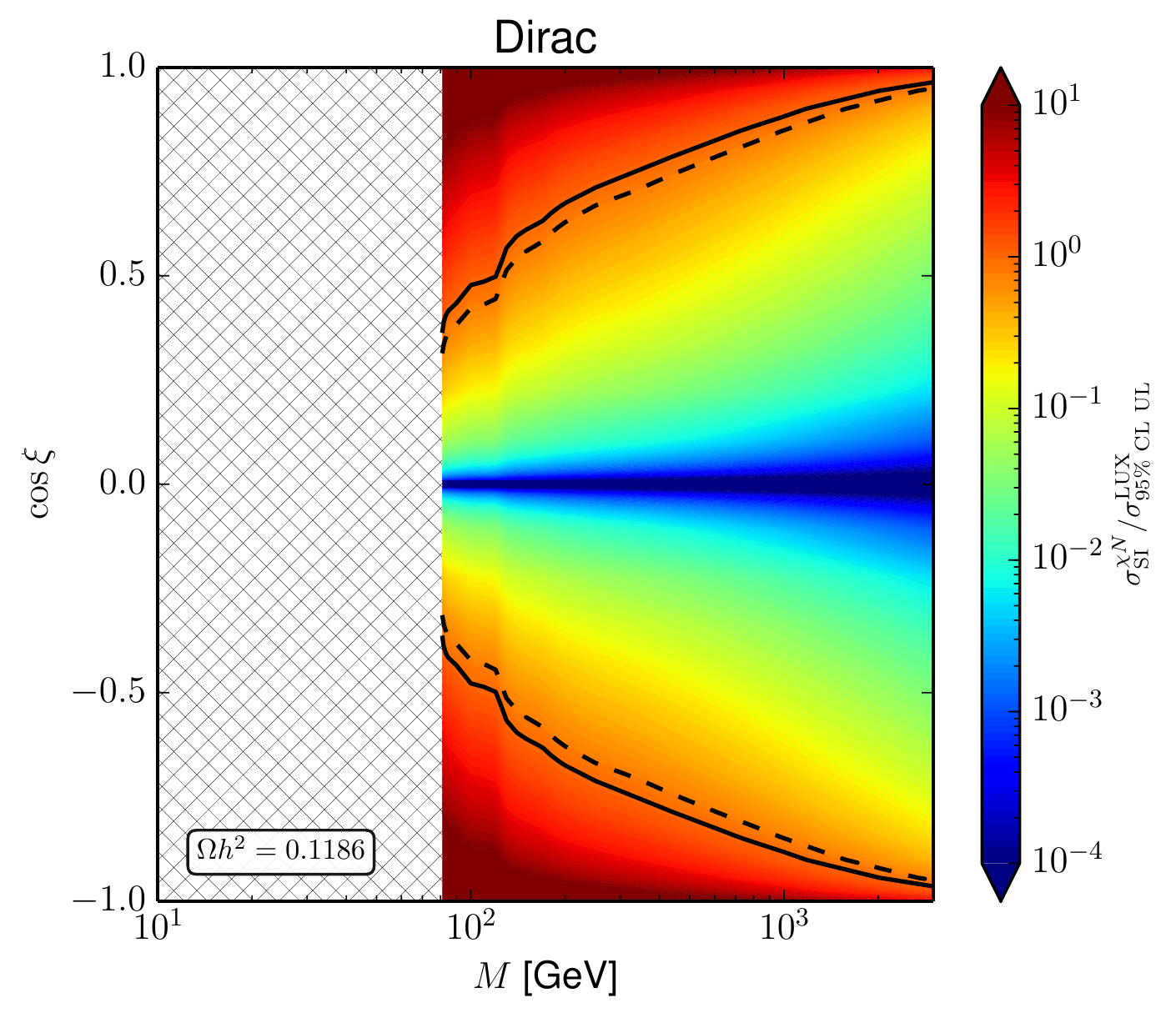}
\includegraphics[width = 0.49 \textwidth]{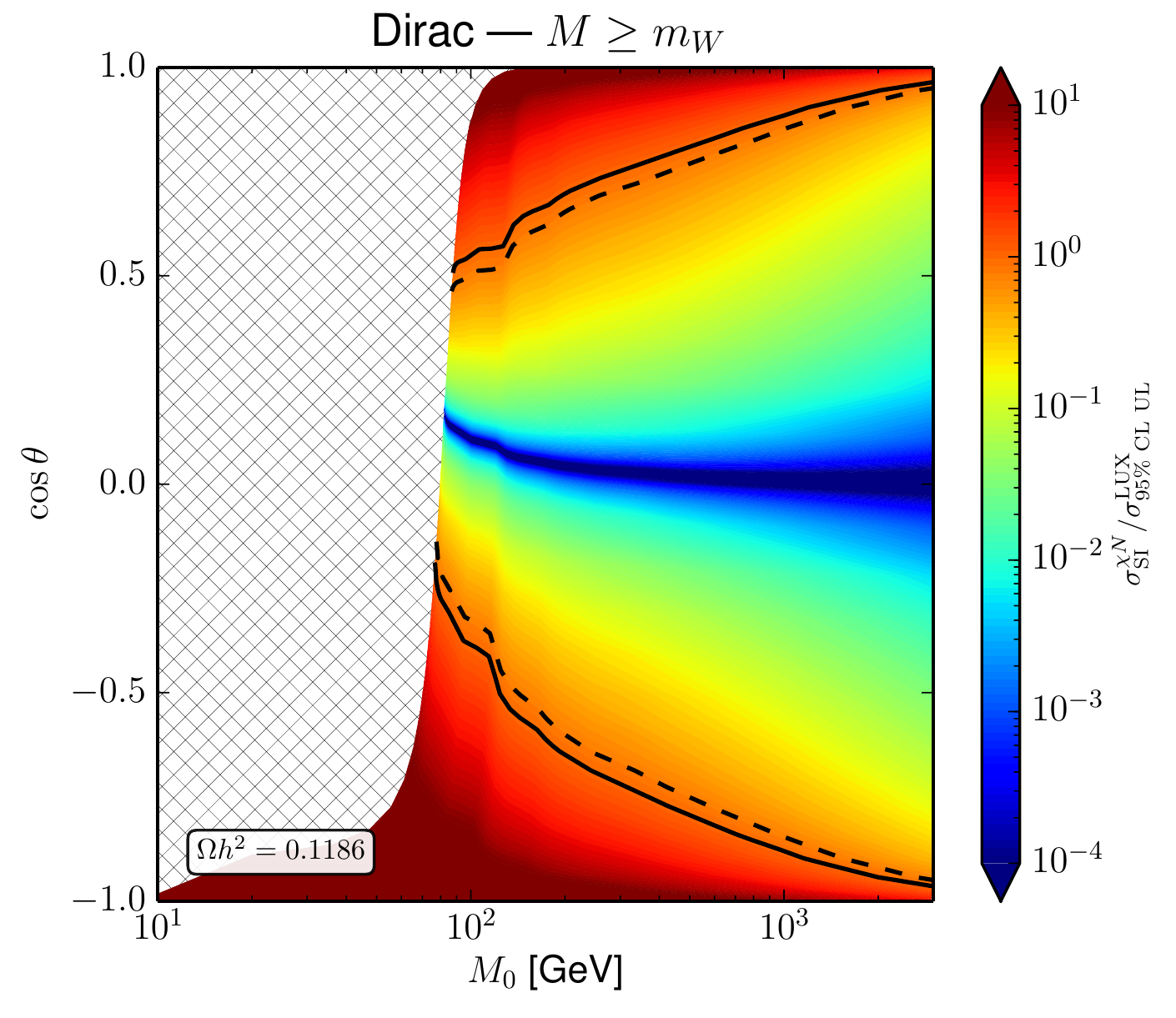} \\
\includegraphics[width = 0.49 \textwidth]{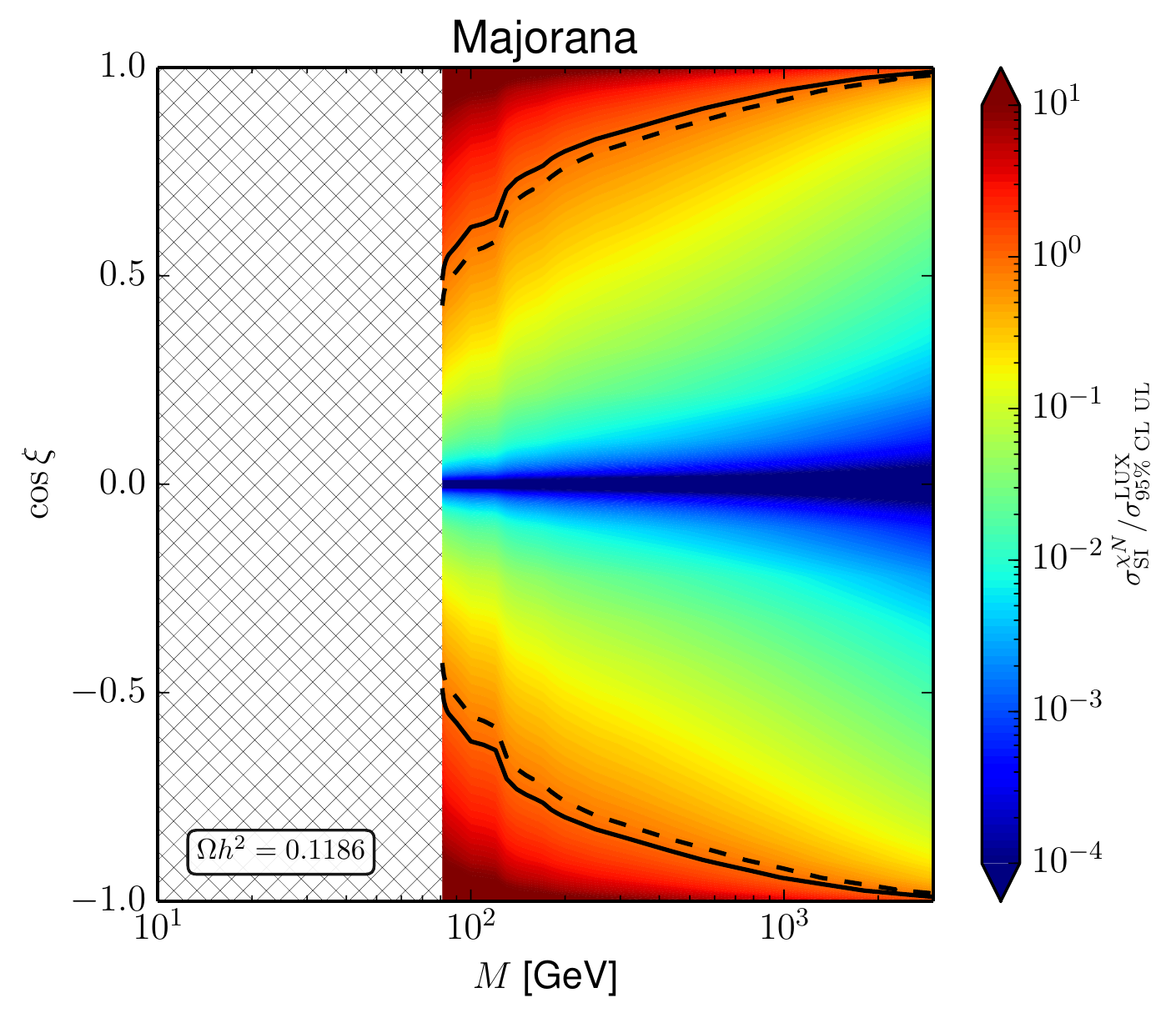}
\includegraphics[width = 0.49 \textwidth]{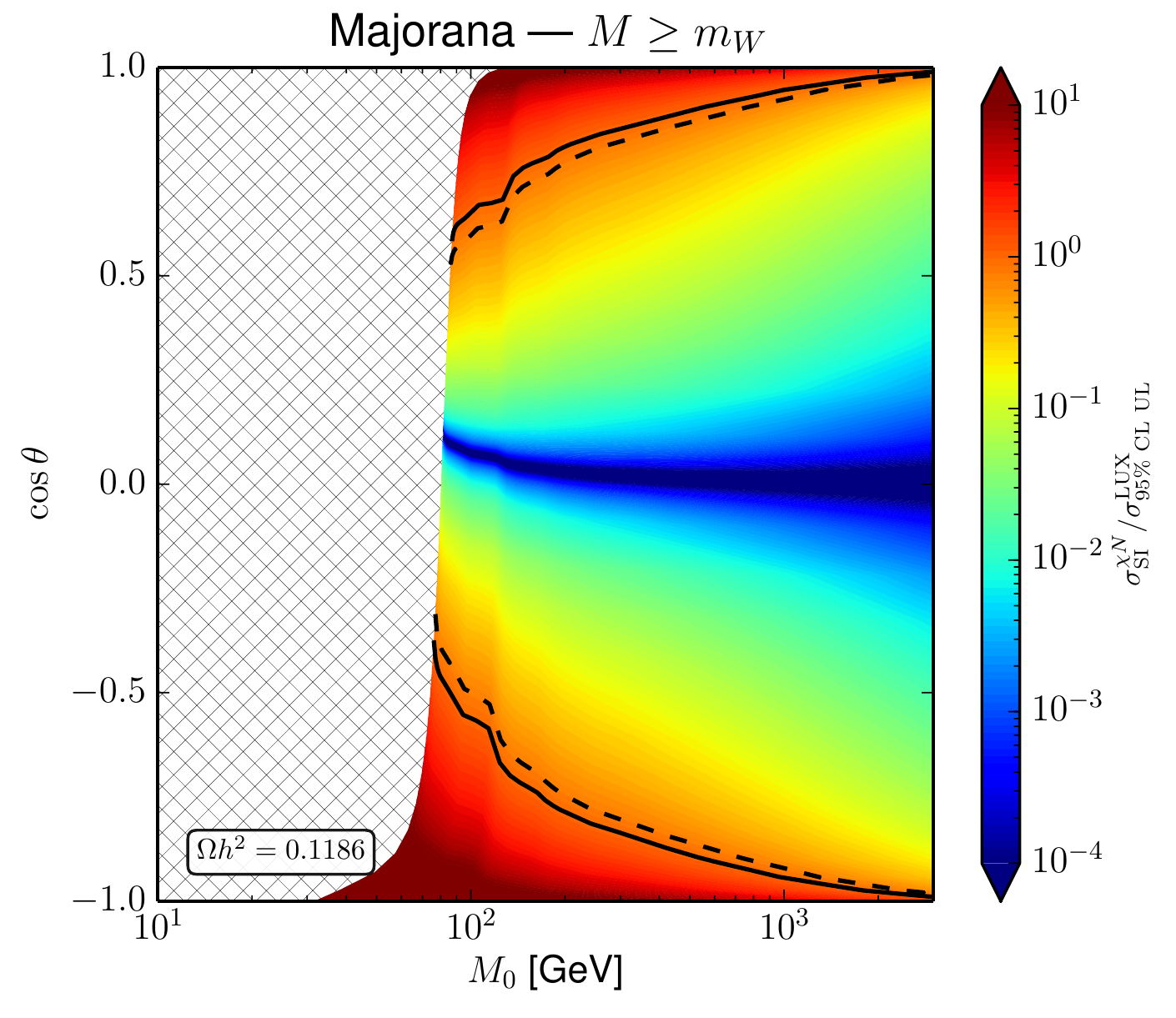}
\end{center}
\caption{\label{app_fig:comp_plot} Direct detection bounds computed in the same way as for figure \ref{fig:direct_exclusions} in the regions where the \emph{physical} DM mass satisfies the constraint $M \geq m_W$, presented in terms of the variables $(M,\cos\xi)$ in the left column and $(M_0,\cos\theta)$ in the right column, for Dirac fermion DM (top row) and Majorana fermion DM (bottom row). All features are as described in the caption of figure \ref{fig:direct_exclusions}, with the exception that the hatching here indicates that $M<M_W$. Important caveats stated in the main text apply to the interpretation of the results in the right column.}
\end{figure}

It is interesting to view the limits we have presented in the main text also in terms of $(M_0,\theta)$, as these are in the parameters which appear in the manifestly gauge-invariant Lagrangian eq.\ \eqref{eq:b4ewsb}. However, as discussed in section \ref{sec:eft}, the map $(M,\xi) \mapsto (M_0,\theta)$ (see eqs.\ \eqref{eq:EMM} and \eqref{eq:cos_sin_xi}) is not necessarily 1-to-1 once we impose the relic density constraint $\Lambda= \Lambda(M,\xi)$. This makes the general presentation of our results in terms of $(M_0,\theta)$ challenging. However provided the \emph{physical} mass of the DM particle, $M$, satisfies the constraint $M\geq m_W$, we can present some restricted results. The form of $\Lambda(M,\xi)$ is sufficiently simple in this region (see figure \ref{fig:DM}) that the map $(M,\xi) \mapsto (M_0, \theta)$ constrained to the domain $M>m_W$ is indeed 1-to-1; or viewed in the other direction, the map $(M_0,\theta) \mapsto (M,\xi)$ is single-valued if restricted to the range $M>m_W$. This allows us to present the results of our analysis of the bounds from direct detection, which are the only ones relevant in the regime $M>m_W$, in terms of the variables $(M_0,\cos\theta)$, which we do in the plots in the right column of figure \ref{app_fig:comp_plot}. We stress that owing to the considerations already outlined, the plots in the right column of figure \ref{app_fig:comp_plot} cannot be interpreted na\"ively as showing regions of $(M_0,\cos\theta)$ which are ruled out by direct detection assuming the correct relic abundance; they may only be interpreted in this fashion if additionally one assumes the constraint on the physical DM mass, $M\geq m_W$. Also shown in the left column of figure \ref{app_fig:comp_plot} are the direct detection constraints in terms of $(M,\cos\xi)$; these results are a subset of those already shown in figure \ref{fig:direct_exclusions} where they were presented as a function of $(M,\cos^2\xi)$. The re-presentation here is to facilitate more direct comparison between the nature of the constrained regions of parameter space when viewed in each set of variables. 

In the large $M$ (or $M_0$) regions the constrained parameter space is broadly similar for the two sets of variables, but as either of these parameters (or $\Lambda$) decreases, the nature of the contained regions begins to differ due to the increasing importance of the chiral rotation (c.f.\ eq.\ \eqref{eq:tan_alpha}): we note in particular that the constraints in terms of $(M_0,\cos\theta)$ (assuming $M\geq m_W$) are not symmetric about $\cos\theta = 0$ whereas those in terms of $(M,\cos\xi)$ are symmetric (provided we continue to ignore the $t$- and $u$- channel diagrams; see figure \ref{fig:HOT}). The origin of this asymmetry is already manifest in eq.\ \eqref{eq:cos_sin_xi}.


\bibliographystyle{JHEP}
\bibliography{fckw}

\end{document}